\documentclass[11pt]{article}
\usepackage{graphicx}
\usepackage{amsmath}
\usepackage{axodraw}
\def\bvg{{B\to V\gamma}}

\def\beq{\begin{equation}}
\def\eeq{\end{equation}}
\def\bea{\begin{eqnarray}}
\def\eea{\end{eqnarray}}
\def\barr{\begin{array}}
\def\earr{\end{array}}

\def\as{\alpha_s}
\def\tp{\tilde{k}_+}

\def\muv{\mu_\mathrm{UV}}
\def\mir{\mu_\mathrm{IR}}
\def\nuv{\nu_\mathrm{UV}}
\def\pv#1{\vec{#1}_\perp}
\def\dirac#1{#1\llap{/}}
\newcommand{\lqcd}{\Lambda_{\textrm{\scriptsize{QCD}}}}

\begin{document}


\begin{titlepage}

\begin{flushright}
LPT-ORSAY 04-27\\
SHEP 04/12\\
\end{flushright}

\vspace{0.3cm}

\begin{center}
{\bf {\Large Spectator Interactions in \boldmath{$B\to V\gamma$} Decays\\
and QCD Factorisation}}

\vspace{1cm}

S.~Descotes-Genon$^a$ and C.T.~Sachrajda$^b$

\vspace{0.5cm}

\emph{$^a$ Laboratoire de Physique Th\'eorique\footnote{LPT is an
Unit\'e Mixte de Recherche du CNRS et de l'Universit\'e Paris XI
(UMR 8627)},
Universit\'e Paris XI}\\
\emph{91405 Orsay Cedex, France}

\vspace{0.2cm}

\emph{$^b$ School of Physics and Astronomy, Univ. of Southampton,}\\
\emph{Southampton, SO17 1BJ, United Kingdom}

\vspace{1.5cm}

{\bf Abstract}
\end{center}

We study the radiative decays $B\to V\gamma$ (where $V=\rho, K^*$)
in the framework of QCD factorisation, and in particular the
hard-spectator contributions to the decay amplitudes. For the
phenomenologically significant chromomagnetic operator, we show by
an explicit next-to-leading-order computation that the spectator
interactions factorise in the heavy-quark limit, i.e. that they
can be written as the convolution of a hard-scattering kernel,
computable in perturbation theory, and of the light-cone
distribution amplitudes for the $B$ and $V$ mesons which contain
the soft physics. The presence of an intermediate scale of
$O(M_B\lqcd)$ leads to the presence of Sudakov logarithms. We
indicate how the demonstration of factorisation can be extended to
other (four-quark) operators.

\end{titlepage}

\section{Introduction}\label{sec:intro}

Radiative $B\to V\gamma$ decays (where $V$ is a light vector
meson) are processes of particular interest in flavour physics
which are already accessible at $B$-factories. Current
measurements yield the following branching
ratios~\cite{Kstar-exp}:
$$B(B^0\to K^{*0}\gamma) = (4.18\pm 0.23)\cdot 10^{-5}
\qquad B(B^+\to K^{*+}\gamma) = (4.14\pm 0.33)\cdot 10^{-5}\,,
$$
whereas only upper bounds are available for the $\omega^0\gamma$,
$\rho^0\gamma$ and $\rho^+\gamma$ modes (of order $O(10^{-6})$ at
90~\% confidence level)~\cite{rho-exp}. Within the standard model
these decays can give measurements of the $V_{td}$ and $V_{ts}$
elements of the CKM matrix.

Penguin-mediated processes, such as $b\to s \gamma$ and $b\to d
\gamma$ radiative decays, exhibit particular sensitivity to
physics beyond the standard model. It is therefore theoretically
important to be able to disentangle the Standard Model
contributions from potential effects of new physics. The principal
difficulty in calculating the amplitudes is due to the presence of
non-pertubative QCD effects. For inclusive decays, such as $B\to
X_s\gamma$, an operator product expansion allows the decay
amplitude to be computed as a series in inverse powers of $m_b$
(the mass of the $b$-quark), with the leading term being
calculable in perturbation theory~\cite{shifman}. For exclusive
decays the decay amplitudes can be calculated in the framework of
QCD factorisation, first developed for nonleptonic two-body
decays~\cite{bbns1,bbns2}. At leading order in $1/M_B$, the
long-distance QCD effects are factorised into universal
quantities, the $B\to$ meson transition form factors and the
light-cone distribution amplitudes of the mesons.

The factorisation framework has been applied to $B\to K^*\gamma$
and $B\to \rho\gamma$ decays, refs.~\cite{BB,BFS}, where in
addition to the leading-twist contributions, the subleading (but
phenomenologically relevant) effects of weak annihilation were
also considered. The same class of decays was also analysed in an
effective-theory framework at higher orders in perturbation
theory~\cite{SCETvgam}. The subsequent phenomenological analysis
suggests interesting quantities to constrain the shape of the
unitarity triangle, such as the $\rho\gamma$ CP-asymmetry for the
$\gamma$ angle~\cite{BB} and the ratio of branching ratios for
$\rho^0\gamma$ and $K^{*0}\gamma$ for the $R_t$ side (which
provides bounds already competitive with the $B-\bar{B}$ and
$B_s-\bar{B}_s$ mass differences)~\cite{Bosch}. These
flavour-changing neutral-current processes yield strong
constraints on supersymmetric models~\cite{ali}.

In view of their phenomenological importance, $B\to V\gamma$
decays deserve further investigation in the framework of QCD
factorisation. In this paper we focus on (strong) radiative
corrections to spectator interactions, in order to determine
whether and how factorisation holds at higher orders in
perturbation theory. We have previously followed a similar line of
investigation for the purely radiative decay
$B\to\gamma\ell\nu$~\cite{dgs_glnu} (and the related decays
$B\to\gamma\gamma$ and
$B\to\gamma\ell^+\ell^-$~\cite{dgs_universal}) in order to acquire
a better understanding of the properties of the light-cone
distribution amplitude of the $B$-meson. $B\to V\gamma$ decays
provide an opportunity to study similar issues concerning the
factorisation of long-distance effects, but now with the presence
of a hadron in the final state. In addition, these decays possess similar
features to heavy-to-light semileptonic decays, such as
$B\to\pi\ell\nu$ which are currently being studied in the SCET
framework~\cite{SCET,SCET2} (see refs.~\cite{beneke2,LN,SCET2-ir} for recent
progress).

The effective Hamiltonian for $B\to V\gamma$ decays, where $V$ is
a light-vector meson $V$ is~\cite{buras}:
\begin{equation}\label{eq:heff}
\mathcal{H}=\frac{G_F}{\sqrt{2}} \sum_{p=u,c} \lambda_p^{(q)}
  \,[C_1 {\mathcal Q}_1^p + C_2 {\mathcal Q}_2^p + \sum_{i=3}^8 C_i {\mathcal Q}_i]
\end{equation}
where $\lambda_p^{(q)}=V^*_{pq}V_{pb}$ (unitarity of the CKM
matrix implies that
$\lambda_t^{(q)}=-(\lambda_u^{(q)}+\lambda_c^{(q)})$ and so
contributions from diagrams with loops containing a top quark are
included implicitly). ${\mathcal Q}_1$ -- ${\mathcal Q}_6$ are
four-quark operators:
\begin{eqnarray}
{\mathcal Q}_1^p=(\bar{q}\,p)_{V-A}\,(\bar{p}\,b)_{V-A}\,&\qquad&
{\mathcal Q}_2^p=(\bar{q}_i\,p_j)_{V-A}\,(\bar{p}_j\,b_i)_{V-A}\\
{\mathcal Q}_3=(\bar{q}\,b)_{V-A}\,\sum_{q^\prime} (\bar{q}^{\,
\prime}\,q^\prime)_{V-A}\,&\qquad& {\mathcal Q}_4=
(\bar{q}_i\,b_j)_{V-A}\,\sum_{q^\prime} (\bar{q}_j^{\,\prime}\,q_i^\prime)_{V-A}\\
{\mathcal Q}_5=(\bar{q}\,b)_{V-A}\,\sum_{q^\prime} (\bar{q}^{\,
\prime}\,q^\prime)_{V+A}\,&\qquad& {\mathcal Q}_6=
(\bar{q}_i\,b_j)_{V-A}\,\sum_{q^\prime} (\bar{q}_j^{\,
\prime}\,q_i^\prime)_{V+A}
\end{eqnarray}
and ${\mathcal Q}_7$ and ${\mathcal Q}_8$ are the electromagnetic
and chromomagnetic penguin operators:
\begin{eqnarray}
{\mathcal Q}_7 = \frac{e}{8\pi^2}m_b\
\bar{q}\,\sigma^{\mu\nu}(1+\gamma_5)b\ F_{\mu\nu}\,&\qquad&
{\mathcal Q}_8 = \frac{g}{8\pi^2}m_b\
\bar{q}\,\sigma^{\mu\nu}(1+\gamma_5)T^A b\ G^A_{\mu\nu}\,.
\end{eqnarray}
$q=d$ or $s$ and the convention for the sign of the coupling
constants corresponds to the covariant derivative
$D_\mu=\partial_\mu+ieQ_fA_\mu+igT^aA_\mu^a$, with $A_\mu$ and
$A_\mu^a$ representing the photon and gluon fields respectively
and $Q_e=-1$ etc. The Wilson coefficients $C_i$ are known at
next-to-leading order in renormalization group improved
perturbation theory~\cite{nlo}.

The calculation of the amplitude for $B\to V\gamma$ decays
requires the evaluation of the matrix elements of the weak
operators ${\mathcal Q}_i$ listed above. We will restrict our
analysis to the phenomenologically relevant operators ${\mathcal
Q}_1,{\mathcal Q}_7,{\mathcal Q}_8$: the QCD penguins ${\mathcal
Q}_3\ldots {\mathcal
  Q}_6$ start at $O(\as)$ and are multiplied by small Wilson
coefficients in the weak Hamiltonian $\mathcal{H}$, whereas the
contribution from ${\mathcal Q_2}$ starts only at $O(\as^2)$.

The factorisation formula for $\bvg$ decays is of the form
\begin{equation}\label{eq:ff}
\frac{G_F}{\sqrt{2}}\, \lambda_p^{(q)}\,\langle
V\gamma\,|\,{\mathcal Q}_i\,|\bar B\rangle=F^{B\to
V}(0)\,T_i^I+\int \frac{d\tp}{2\pi}\,du\
\Phi^B(\tp)\,T^{II}_i(\tp,u)\Phi_\perp(u)\,,
\end{equation}
where the non-perturbative effects are contained in $F^{B\to V}$,
a form factor for $B\to V$ transitions, and $\Phi^B$ and
$\Phi_\perp$, the leading-twist light-cone distribution amplitudes
of the $B$ and $V$ mesons. The hard-scattering amplitudes $T^I_i$
and $T^{II}_i$ include only short-distance effects and are
calculable in perturbation theory. The indices $p$ and $q$ have
been suppressed on the right-hand side of eq.~(\ref{eq:ff}). It is
expected that the factorisation formula is valid up to corrections
of $O(\lqcd/m_b)$. Following the analysis in refs.~\cite{BB,BFS},
the contribution from the electromagnetic operator ${\mathcal
Q}_7$ is included in the first term on the right-hand side of
eq.~(\ref{eq:ff}), corresponding to a form factor. The
chromomagnetic operator ${\mathcal Q}_8$ and the four-quark
operator ${\mathcal Q}_1$ contribute to both terms.

The subject of this paper is the second term on the right-hand
side of eq.~(\ref{eq:ff}), which arises from hard spectator
interactions. For such terms factorisation has been explicitly
established only at leading order in perturbation theory. We
consider the spectator contribution to the matrix element of the
chromomagnetic operator ${\mathcal Q}_8$, and demonstrate by
explicit calculation that the mass singularities at one-loop order
are precisely those of the distribution amplitudes in
eq.~(\ref{eq:ff}). The detailed results for the mass singularities
from each diagram are presented in appendix~\ref{app:explicit},
together with the terms containing ``large logarithms". We analyse
the reasons for the factorisation of mass singularities, and
present a heuristic demonstration for the matrix elements of
${\mathcal Q}_8$ in section~\ref{sec:fact1lpo8}. We then argue
that such a cancellation is also valid for the matrix elements of
the remaining operators (appendix~\ref{app:general}), analysing
the contribution from ${\mathcal Q}_1$, which is expected to be
the largest from the four-quark operators ${\mathcal
Q}_1$\,--\,${\mathcal Q}_6$, in particular detail in
section~\ref{sec:fact1lpo1}.
Section~\ref{sec:concs} contains our conclusions.

\subsection{Kinematics}

We consider the decay
\begin{equation}
B(p)\to \gamma(q_1,\varepsilon^\ast) V(q_2,\eta^*)\,
\end{equation}
where $p,\,q_1$ and $q_2$ are the momenta of the corresponding
particles and $\varepsilon^\ast$ and $\eta^\ast$ are the
polarization vectors of the photon and $V$ respectively. We work
in the rest frame of the $B$-meson, $p=(M_B,\vec 0)$, and take the
momenta of the photon and $V$ to be in the $z$-direction, with
both final-state vectors transversely polarized.

It is convenient to introduce light-cone coordinates; for any
four-vector $l$ we write $l=(l_+,l_-,\pv{l})$, with
\begin{equation}
l_\pm = \frac{l_0\pm l_3}{\sqrt{2}}\qquad\textrm{so that}\qquad
l^2=2l_+l_--\pv{l}^{\hspace{2.5pt}2}\,.
\end{equation}
We neglect the mass of the vector meson, and take $q_1$ and $q_2$
to be in the $+$ and $-$ directions respectively:
\begin{equation}
q_1=\left(\frac{M_B}{\sqrt{2}},0,\pv{0}\right),\quad
q_2=\left(0,\frac{M_B}{\sqrt{2}},\pv{0}\right)\,.
\end{equation}

\subsection{Evaluating the \boldmath{$T^{II}_i$}}

In this paper we focus on the contribution from the hard spectator
interactions represented by the second term on the right-hand side
of eq.~(\ref{eq:ff}), which we write in the schematic form
$\Phi^B\otimes T_i\otimes \Phi_\perp$. Here and in the remainder
of the paper, for simplicity of notation, we suppress the
superscript {\small${II}$}.

In order to verify the validity of the factorisation
formula~(\ref{eq:ff}) we can use perturbation theory with
conveniently chosen partonic external states. In particular we
evaluate the hard-scattering kernels $T_i$, and a signal of the
breakdown of factorisation would be the presence of residual
long-distance effects in the $T_i$. For our calculations we take
the initial state to have momentum $p$ and to consist of a
$b$-quark with momentum $p-k$ and a light (spectator) antiquark
$\bar{q}^\prime$ with momentum $k$. The components of $k$ are of
$O(\lqcd)$. Our final state consists of the photon with momentum
$q_1$ and a quark ($q$) and spectator antiquark
($\bar{q}^{\,\prime}$) with momenta $\bar{x}q_2$ and $xq_2$
respectively. The partons in the initial and final states are on
their mass shells. The hard-scattering kernel is obtained in the
standard way; by using perturbation theory to calculate the matrix
elements of the operator ${\cal Q}_i$ between these states
together with the light-cone distribution amplitudes (defined
explicitly in the next subsection) for the initial and final
partonic states. Let ${\cal A}_i$ denote the left hand side of
eq.\,(\ref{eq:ff}) and the superscript {\small($n$)} label the
perturbative contributions of $O(\alpha_s^n)$ relative to the
lowest order (denoted by {\small($0$)}). The hard-scattering
kernels at one-loop order ($T_i^{(1)}$) are then obtained by
rewriting the factorisation formula as
\begin{equation}\label{eq:t1def}
\Phi^{b\bar{q}^\prime\,(0)} \otimes T_i^{(1)} \otimes
\Phi^{q\bar{q}^\prime\,(0)}= {\cal
A}_i^{(1)}-\Phi^{b\bar{q}^\prime\,(1)} \otimes T_i^{(0)} \otimes
\Phi^{q\bar{q}^\prime\,(0)}-\Phi^{b\bar{q}^\prime\,(0)} \otimes
T_i^{(0)} \otimes \Phi^{q\bar{q}^\prime\,(1)}\,,
\end{equation}
where the $\Phi$'s denote the corresponding distribution
amplitudes. We confirm below that although there are mass
singularities in the ${\cal A}_i^{(1)}$, they are cancelled by the
remaining terms on the right-hand side of eq.~(\ref{eq:t1def}).

Throughout this paper the discussion of the contributions from
particular diagrams corresponds to the Feynman gauge.

\subsection{Light-Cone Distribution Amplitudes}\label{subsec:lcda}

We define the light-cone distribution amplitude of a state $H$
containing the $b$-quark by
\begin{equation}\label{eq:phihdef}
\Phi^H_{\alpha\beta}(\tp)=\int\, dz_-\ e^{i\tp z_-} \langle 0|
\bar{q}^{\,\prime}_\beta(z)[z,0]
b_\alpha(0)|H\rangle|_{z_+,z_\perp=0}\,,
\end{equation}
where $q^{\,\prime},b$ are the quark fields and $\alpha,\beta$ are
spinor labels (we denote the spectator antiquark by
$\bar{q}^{\,\prime}$). $[z,0]$ denotes the path-ordered
exponential, $\mathcal{P}\exp[-ig\int_0^z dx^\mu A_\mu(x)]$.
Ultimately of course, when evaluating the physical decay amplitude
we use the distribution amplitude of the $B$-meson, but in the
evaluation of the hard-scattering amplitude $T_i$ in perturbation
theory we take $H$ to be a quark-antiquark state.

Similarly the light-cone distribution amplitude of a light state
$L$ is defined by:
\begin{equation}\label{eq:phimdef}
\Phi^L_{\gamma\delta}(u)=q_{2-}
  \int \frac{d(y-x)_+}{2\pi}\ e^{-i(uq_2\cdot x+\bar{u}q_2\cdot y)}\
  \langle L\,|\, \bar{q}_\delta(y)[y,x] q^{\,\prime}_\gamma(x)
            \,|\,0\rangle|_{(x-y)_-,(x-y)_\perp=0}\ ,
\end{equation}
where $\bar{u}=1-u$, and the integrand in eq.~(\ref{eq:phimdef})
is a function of $(x-y)_+$.

In terms of the definitions in equations~(\ref{eq:phihdef}) and
(\ref{eq:phimdef}), for the $B$ and $V$ mesons we follow an
equivalent spinor decomposition to that in
refs.~\cite{bf,bbns2,BB} and define the distribution amplitudes
$\Phi^B_\pm(\tp)$ and $\Phi_\perp(u)$ by~\footnote{Following
ref.~\cite{dgs_glnu,dgs_universal}, $\Phi^B_\pm$ is defined in the
Heavy-Quark Effective Theory, but with the physical decay constant
$f_B$ in front of it. It was pointed out in ref.~\cite{neubertrad}
that a different normalisation factor can be chosen to ensure a
more natural comparison with HQET results. Such a change in the
normalisation of $\Phi^B_\pm$ does not affect the outcome of our
analysis and corresponds to a simple redistribution of terms
between the hard-scattering kernel $T^{II}$ and the $B$-meson's
light-cone distribution amplitude.}
\begin{eqnarray}
\Phi^B_{\alpha\beta}(\tp)&=&-\frac{if_BM_B}{4}
  \left\{(1+\dirac{v})
       \left[\Phi_+^B(\tp)+\dirac{n}_- (\Phi_-^B(\tp)-\Phi_+^B(\tp))\right]
\gamma_5 \right\}_{\alpha\beta}\\
\Phi^V_{\gamma\delta}(u)&=&-\frac{if_V^\perp}{4}
  (\sigma_{\mu\nu})_{\gamma\delta}\ (\eta^*)^\mu q_2^\nu
  \Phi_\perp(u)=\frac{f_V^\perp}{4}\dirac{\eta}^\ast\dirac{q}_2\Phi_\perp(u)
  \,,\label{eq:phiperpdef}
\end{eqnarray}
where the light-like vector $n_-$ is given in light-cone
coordinates by $n_-=(0,\sqrt{2},\pv{0})$, and the four velocity
$v$ is defined by $p_\mu=M_Bv_\mu$. The decay constants $f_B$ and
$f_V^\perp$ are defined by
\begin{equation}
\langle
0\,|\bar{q}^{\,\prime}\gamma_\mu\gamma_5b\,|\,\overline{B}(p)\rangle=
if_Bp_\mu\qquad\textrm{and}\qquad \langle
V(k,\eta^\ast)\,|\bar{q}\sigma_{\mu\nu}q^{\,\prime}\,|\,0\rangle=
-i(\eta^\ast_\mu k_\nu-\eta^\ast_\nu k_\mu)\,f_V^\perp\,.
\end{equation}

\section{Leading Order Results for the Matrix Elements of the
Operators \boldmath{${\cal Q}_8$} and \boldmath{${\cal Q}_1$}}
\label{sec:lowest}

Before studying the cancellation of mass singularities at
next-to-leading order (NLO) in perturbation theory, we present the
lowest-order contributions to the hard-scattering amplitude from
the operators ${\cal Q}_8$ and ${\cal Q}_1$ respectively.

\subsection{Tree-Level Result for the matrix element of \boldmath{${\cal
Q}_8$}}

As explained above, we determine the hard-scattering amplitude
$T_8$ by choosing the convenient external states
$H=b\bar{q}^{\,\prime}$ and $L=q\bar{q}^{\,\prime}$ and computing
both the matrix element
\begin{equation}\label{eq:fhmdef}
{\cal A}_8\equiv -\frac{G_F\lambda^{(q)}_t}{\sqrt{2}}
     \langle L\gamma|{\cal Q}_8|H\rangle
\end{equation}
and the distribution amplitudes $\Phi^{H,L}$. In this section the
calculation is performed at lowest order of perturbation theory.

There are four diagrams contributing to ${\cal A}_8$ at lowest
order, and these are shown in fig.~\ref{fig:tree8}. At leading
twist only the diagram of fig.~\ref{fig:tree8}(a) contributes and
we find:
\begin{equation}
{\cal A}^{(0)}_8=-G_F \lambda_t^{(q)}\,\frac{eQ_{q}\as}{2\pi}\,
    \frac{1}{\bar{x}k_+}\ \Big(\bar{u}(\bar{x}q_2)
   \left\{\dirac\epsilon^*\gamma_\nu(1+\gamma_5)T^A\right\}u(p-k)\Big)
    \ \Big(\bar{v}(k)\left\{\gamma^\nu T^A\right\}v(xq_2)\Big)\,,
\end{equation}
where $u$ and $v$ represent the free-particle spinor wave
functions and $Q_{q}=-1/3$ is the charge of the quark $q$.

\begin{center}
\begin{figure}[t]
\begin{picture}(300,230)(-70,-180)
\ArrowLine(0,0)(50,0)\ArrowLine(50,0)(100,0)
\SetWidth{1.5}\Line(0,60)(50,60)\SetWidth{0.5} \Line(50,60)(80,60)
\ArrowLine(80,60)(100,60) \ArrowLine(24,60)(26,60)
\Photon(50,60)(50,0){2}{10} \Gluon(65,60)(90,40){2}{4}
\Text(25,-4)[t]{\small{$k$}}\Text(75,-4)[t]{\small{$xq_2$}}
\Text(25,64)[b]{\small{$p-k$}}\Text(95,64)[b]{\small{$\bar{x}q_2$}}
\Text(94,40)[l]{\small{$q_1$}}\GBox(45,55)(55,65){1}
\Text(50,-20)[t]{\small{(a)}}
\ArrowLine(200,0)(250,0)\ArrowLine(280,0)(300,0)\Line(250,0)(280,0)
\SetWidth{1.5}\Line(200,60)(250,60)\SetWidth{0.5}
\ArrowLine(250,60)(300,60) \ArrowLine(224,60)(226,60)
\Photon(250,60)(250,0){2}{10} \Gluon(265,0)(290,20){2}{4}
\Text(225,-4)[t]{\small{$k$}}\Text(295,-4)[t]{\small{$xq_2$}}
\Text(225,64)[b]{\small{$p-k$}}\Text(275,64)[b]{\small{$\bar{x}q_2$}}
\Text(294,20)[l]{\small{$q_1$}}\GBox(245,55)(255,65){1}
\Text(250,-20)[t]{\small{(b)}}
\ArrowLine(0,-150)(50,-150)\ArrowLine(50,-150)(100,-150)
\SetWidth{1.5}\Line(0,-90)(50,-90)\SetWidth{0.5}
\Line(50,-90)(80,-90) \ArrowLine(80,-90)(100,-90)
\ArrowLine(24,-90)(26,-90) \Photon(50,-90)(50,-150){2}{10}
\Gluon(5,-150)(30,-130){2}{4}
\Text(25,-154)[t]{\small{$k$}}\Text(75,-154)[t]{\small{$xq_2$}}
\Text(25,-86)[b]{\small{$p-k$}}\Text(95,-86)[b]{\small{$\bar{x}q_2$}}
\Text(34,-130)[l]{\small{$q_1$}}\GBox(45,-95)(55,-85){1}
\Text(50,-170)[t]{\small{(c)}}
\ArrowLine(200,-150)(250,-150)\ArrowLine(280,-150)(300,-150)\Line(250,-150)(280,-150)
\SetWidth{1.5}\Line(200,-90)(250,-90)\SetWidth{0.5}
\ArrowLine(250,-90)(300,-90) \ArrowLine(224,-90)(226,-90)
\Photon(250,-90)(250,-150){2}{10} \Gluon(205,-90)(230,-110){2}{4}
\Text(225,-154)[t]{\small{$k$}}\Text(295,-154)[t]{\small{$xq_2$}}
\Text(225,-86)[b]{\small{$p-k$}}\Text(275,-86)[b]{\small{$\bar{x}q_2$}}
\Text(234,-110)[l]{\small{$q_1$}}\GBox(245,-95)(255,-85){1}
\Text(250,-170)[t]{\small{(d)}}
\end{picture}
\caption{Lowest-order contributions to the matrix element $\langle
\bar{q}^{\,\prime}(xq_2)\,q(\bar{x}q_2)\gamma(q_1)\, |\,{\cal
Q}_8\,|\,\bar{q}^{\,\prime}(k)\,b(p-k)\rangle$. The spring-like
line represents the photon and the square denotes the insertion of
${\cal Q}_8$.}\label{fig:tree8}
\end{figure}
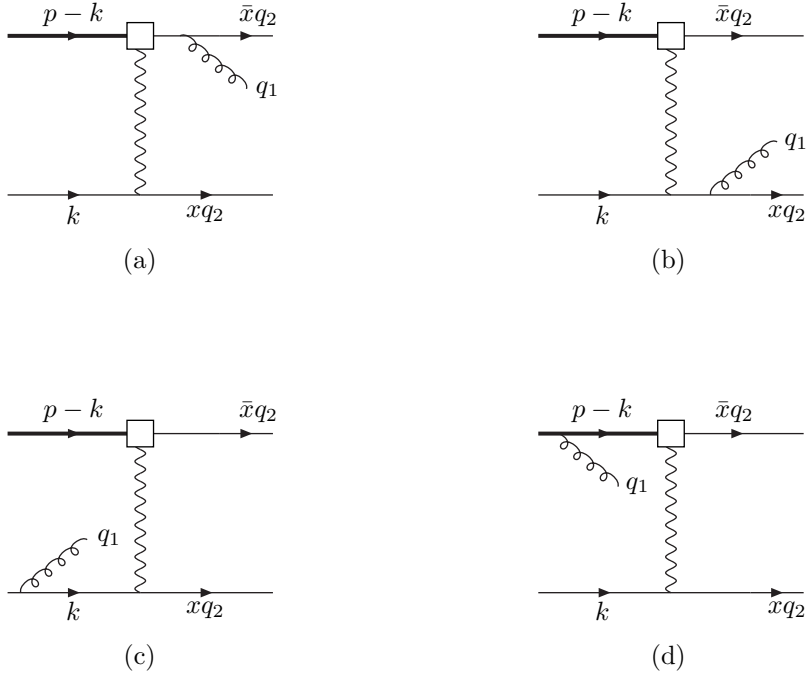
\end{center}\vspace{-40pt}

The distribution amplitudes at lowest order are:
\begin{eqnarray}\label{eq:phih0}
\Phi^{b\bar{q}^\prime\,(0)}_{\alpha\beta}(\tp)&=&
2\pi\delta(k_+-\tp) \ \bar{v}_\beta(k)
    u_\alpha(p-k)\\
\Phi^{q\bar{q}^\prime\,(0)}_{\gamma\delta}(u)&=& \delta(u-x)\
  \bar{u}_\delta(\bar{x}q_2) v_\gamma(xq_2)\,,\label{eq:phim0}
\end{eqnarray}
and ${\cal A}_8^{(0)}$ can be written in the factorised form
\begin{equation}
{\cal A}^{(0)}_8=\Phi^{b\bar{q}^\prime\,(0)}\otimes
T_8^{(0)}\otimes \Phi^{q\bar{q}^\prime\,(0)}
  \end{equation}
with
\begin{eqnarray}\label{eq:t80}
T^{(0)}_{8\ \alpha\beta\gamma\delta}(\tp,u)=
   -G_F \lambda_t^{(q)}\,\frac{eQ_{q}\as}{2\pi}\,
    \frac{1}{\bar{u}\tp}\ \left\{
     \dirac\epsilon^*\gamma_\nu(1+\gamma_5)T^A\right\}_{\delta\alpha}\
     \left\{\gamma^\nu T^A\right\}_{\beta\gamma}\,.
\end{eqnarray}
For simplicity of notation, we have suppressed the colour indices.
$T_8^{(0)}$ is a function of the convolution variables $\tp$ and
$u$, but does not depend on any kinematical quantities of
$O(\lqcd)$.

We conclude the lowest-order calculation by briefly considering
the three diagrams in fig.~\ref{fig:tree8}(b)\,--\,(d).
Fig.~\ref{fig:tree8}(b) manifestly gives a higher-twist
contribution, since both the gluon and internal antiquark are
off-shell by an amount of $O(M_B^2)$. In diagram
\ref{fig:tree8}(c), although the gluon is off-shell by $O(M_B^2)$,
the internal antiquark is off-shell by
$O(\,(q_1-k)^2)=O(M_B\lqcd)$. However this potentially
leading-twist contribution vanishes when projected with the wave
function of a transversely polarized vector meson (see
eq.~(\ref{eq:phiperpdef})\,). The contribution from diagram
\ref{fig:tree8}(d) vanishes by the equations of motion of the
final-state meson.

\subsection{Lowest-order result for the matrix element of
\boldmath{${\cal Q}_1$}}

\begin{center}
\begin{figure}
\begin{picture}(300,80)(-70,-10)
\ArrowLine(0,0)(50,0)\ArrowLine(50,0)(100,0)
\SetWidth{1.5}\Line(0,60)(50,60)\SetWidth{0.5}
\ArrowLine(50,60)(100,60)
\ArrowLine(24,60)(26,60)\GCirc(50,60){2}{0}
\GCirc(50,55){2}{0}\Oval(50,40)(15,8)(0)\Photon(50,25)(50,0){2}{4}
\ArrowLine(42,39.5)(42,40.5)\Gluon(58,40)(90,40){2}{4}
\Text(25,-4)[t]{\small{$k$}}\Text(75,-4)[t]{\small{$xq_2$}}
\Text(25,64)[b]{\small{$p-k$}}\Text(75,64)[b]{\small{$\bar{x}q_2$}}
\Text(94,40)[l]{\small{$q_1$}}
\ArrowLine(200,0)(250,0)\ArrowLine(250,0)(300,0)
\SetWidth{1.5}\Line(200,60)(250,60)\SetWidth{0.5}
\ArrowLine(250,60)(300,60)
\ArrowLine(224,60)(226,60)\GCirc(250,60){2}{0}
\GCirc(250,55){2}{0}\Oval(250,40)(15,8)(0)\Photon(250,25)(250,0){2}{4}
\ArrowLine(242,40.5)(242,39.5)\Gluon(258,40)(290,40){2}{4}
\Text(225,-4)[t]{\small{$k$}}\Text(275,-4)[t]{\small{$xq_2$}}
\Text(225,64)[b]{\small{$p-k$}}\Text(275,64)[b]{\small{$\bar{x}q_2$}}
\Text(294,40)[l]{\small{$q_1$}}
\end{picture}
\caption{Lowest-order contribution to the matrix element $\langle
\bar{q}^{\,\prime}(xq_2)\,q(\bar{x}q_2)\,|\,{\cal
Q}_1\,|\,\bar{q}^{\,\prime}(k)\,b(p-k)\rangle$. The arrow on the
quark loop defines the direction of the flow of fermion
number.}\label{fig:o1lo}
\end{figure}
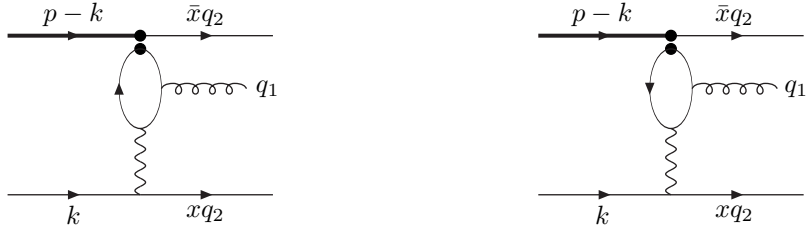
\end{center}\vspace{-35pt}

The second example which we study in some detail is the matrix
element of the operator ${\cal Q}_1$. At lowest order the relevant
diagrams are shown in fig.~\ref{fig:o1lo}. The contribution
proportional to $\lambda_c^{(q)}$ ($\lambda_u^{(q)}$) is the
difference of the diagrams with the charm (up) quark propagating
in the loop and one with the top quark. Here we present the
results for the charm quark (the corresponding results for the up
quark can readily be deduced from these):
\begin{equation}\label{eq:fa10}
{\cal A}_1\equiv \frac{G_F\lambda^{(q)}_c}{\sqrt{2}}
     \langle
     L\gamma|\,(\bar{q}c)_{V-A}\,(\bar{c}b)_{V-A}\,|H\rangle\,.
\end{equation}
Following ref.~\cite{BB}, we do not neglect $m_c^2/m_b^2$, and
hence keep an important imaginary part which vanishes when $m_c$
is set to zero.

Evaluating the diagrams in fig.~\ref{fig:o1lo} we find:
\begin{eqnarray}\label{eq:a10}
{\cal A}_1^{(0)}&=&i\,\frac{G_F\lambda^{(q)}_c}{\sqrt{2}}\
\frac{eQ_c\alpha_s}{4\pi}\frac{2}{xq_{2\,-}k_+}\
\Big(\bar{u}(\bar{x}q_2)\gamma_\nu(1-\gamma_5) T^au(p-k)\Big)\
\Big(\bar{v}(k)\gamma_\rho T^a v(xq_2)\Big)\times\nonumber\\
&&\hspace{-0.6in}\left\{
T_1(x,s)\epsilon^{\sigma\rho\lambda\nu}(xq_2-q_1)_\sigma +
T_2(x,s)\epsilon^{\sigma\lambda\tau\nu}q^\rho_{1}q_{1\,\sigma}q_{2\,\tau}
+
T_3(x,s)\epsilon^{\sigma\rho\tau\lambda}q^\nu_{1}q_{1\,\sigma}q_{2\,\tau}
\right\}\times\varepsilon^\ast_\lambda\,,\end{eqnarray}
where~$Q_c=2/3$ is the charge of the charm quark, $s=m_c^2/m_b^2$,
\begin{eqnarray}
T_1(x,s)&=&-\frac13+\frac{4s}{x}-\frac{2s}{x}L,\label{eq:t1}\\
T_2(x,s)&=&\frac{1}{M_B^2}\left(\frac23-\frac{8s}{x}+\frac{4s}{x}L\right),\label{eq:t2}\\
T_3(x,s)&=&-\frac{1}{M_B^2}\left(\frac43+\frac{8s}{x}-\frac{4s}{x}(L+P)\right),\label{eq:t3}
\end{eqnarray}
with
\begin{equation}
L=\sqrt{1-4s/x+i\epsilon}\,\log\left(\frac{\sqrt{1-4s/x+i\epsilon}+1}
{\sqrt{1-4s/x+i\epsilon}-1}\right)
\end{equation}
and
\begin{equation}
P=L_2\left(\frac{2}{1-\sqrt{\frac{x-4s+i\epsilon}{x}}}\right)
+L_2\left(\frac{2}{1+\sqrt{\frac{x-4s+i\epsilon}{x}}}\right)\,,
\end{equation}
where $L_2$ is the dilogarithm function. From eq.~(\ref{eq:a10})
and the tree-level expressions for the distribution amplitudes in
equations (\ref{eq:phih0}) and (\ref{eq:phim0}) one immediately
obtains the lowest-order contribution to the hard-scattering
kernel $T_1^{(0)}$.

\subsection{Predictions for the Spectator Contributions to \boldmath{$B\to
V\gamma$} Decay Amplitudes at Lowest Order}

In order to obtain the predictions for the spectator contributions
to $B\to V\gamma$ decay amplitudes, the hard-scattering kernels
obtained in the previous two subsections need to be convoluted
with the $B$ and $V$ distribution amplitudes. Convoluting the
lowest-order result for the hard-scattering kernel $T_8^{(0)}$ in
eq.~(\ref{eq:t80}) with the distribution amplitudes we obtain for
the spectator contribution to the amplitude
\begin{eqnarray}
-\frac{G_F\lambda_t^{(q)}}{\sqrt{2}}\langle V\gamma|{\cal
Q}_8|\bar{B}\rangle&=&-(G_F\,\lambda_t^{(q)})\,\frac{\alpha_sC_F}{4\pi}
\,\frac{f_Bf_V^\perp}{N}\,(eQ_q)\ \int\frac{d\tp}{2\pi}\,
\frac{\Phi^B_+(\tp)}{\tp}\,
\int\,du\frac{\Phi_V^\perp(u)}{\bar{u}}\nonumber\\
&&\times\ \left\{\varepsilon_{\mu\nu\lambda\rho}\,
\varepsilon^{\ast\,\mu}\eta^{\ast\,\nu}q_2^\lambda
q_1^\rho+i(\varepsilon^\ast\cdot\eta^\ast)(q_1\cdot
q_2)\right\}\,,
\end{eqnarray}
where $Q_q=-1/3$ is the charge of the down or strange quark and
$N=3$ is the number of colours. This is equivalent to the result
in eqs.~(39) and (40) of ref.~\cite{BB}.

Similarly, convoluting the hard-scattering kernel $T_1^{(0)}$ from
the lowest-order amplitude in eq.~(\ref{eq:a10}) with the $B$ and
$V$-meson distribution amplitudes we obtain
\begin{eqnarray}
\frac{G_F\lambda_c^{(q)}}{\sqrt{2}}\langle V\gamma|{\cal
Q}_1^c|\bar{B}\rangle&=&\frac{G_F\lambda_c^{(q)}}{4}
\,\frac{\alpha_sC_F}{4\pi} \,\frac{f_Bf_V^\perp}{N}\,(eQ_u)\
\int\frac{d\tp}{2\pi}\, \frac{\Phi^B_+(\tp)}{\tp}\,
\int\,du\,\Phi_V^\perp(u)\,h(u,s)\nonumber\\
&&\times\ \left\{\varepsilon_{\mu\nu\lambda\rho}\,
\varepsilon^{\ast\,\mu}\eta^{\ast\,\nu}q_2^\lambda
q_1^\rho+i(\varepsilon^\ast\cdot\eta^\ast)(q_1\cdot
q_2)\right\}\,,\label{eq:a11}
\end{eqnarray}
where
\begin{equation}
h(u,s)=-\frac2u+\frac{4s}{u^2}\left\{L_2\left(\frac{2}{1-\sqrt{\frac{u-4s+i\epsilon}{u}}}\right)
+L_2\left(\frac{2}{1+\sqrt{\frac{u-4s+i\epsilon}{u}}}\right)\right\}\,.
\end{equation}
If $m_c$ is set to zero, $h(u,0)=-2/u$. These results are
equivalent to those in eqs.~(33)\,--\,(35) of ref.~\cite{BB}.

\section{Factorisation at One-Loop Order for the Chromomagnetic
Operator \boldmath{${\cal Q}_8$}}\label{sec:fact1lpo8}

\begin{figure}[t]
\begin{center}
\begin{picture}(230,150)(-15,-5)
\SetWidth{1.5}\Line(0,100)(80,100)\SetWidth{0.8}
\ArrowLine(9,100)(10,100)\SetWidth{0.5}
\Text(-5,100)[r]{\small{$p$\,--\,$k$}}
\GCirc(40,100){7}{1}\Text(40,100)[]{1}
\Line(80,100)(200,100)\ArrowLine(190,100)(191,100)
\GCirc(170,100){7}{1}\Text(170,100)[]{2}
\GCirc(110,100){7}{1}\Text(110,100)[]{5}
\Gluon(140,100)(170,130){2.5}{4}\Text(175,130)[l]{\small{$q_1$}}
\ArrowLine(170,130)(171,131)\Text(205,100)[l]{\small{$\bar{x}q_2$}}
\Photon(80,100)(80,0){3}{14}
\GCirc(80,50){7}{1}\Text(80,50)[]{6}
\Line(0,0)(200,0)\ArrowLine(9,0)(10,0)\ArrowLine(190,0)(191,0)
\Text(205,0)[l]{\small{$xq_2$}}
\Text(-5,0)[r]{\small{$k$}}
\GCirc(40,0){7}{1}\Text(40,0)[]{3}
\GCirc(140,0){7}{1}\Text(140,0)[]{4}
\GBox(70,90)(90,110){1}
\GCirc(80,100){7}{1}\Text(80,100)[]{7}
\end{picture}
\caption{Notation for one-loop diagrams contributing to the $\bvg$
decay from ${\cal Q}_8$. The extra gluon can be attached to any
pair of circles (which might be identical).} \label{fig:1loop8}
\end{center}
\end{figure}
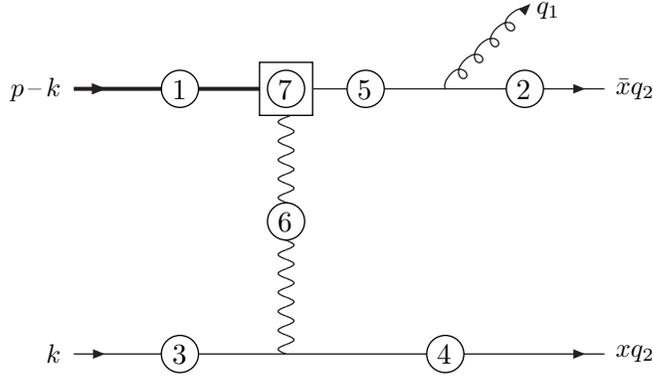

We now turn to the main subject of this paper, the factorisation
of mass singularities at one-loop order for the spectator
interactions. We start with the operator ${\cal Q}_8$ and study
the one-loop corrections to the lowest-order diagram in
fig.~\ref{fig:tree8}(a). We redraw this diagram in
fig.~\ref{fig:1loop8}, labelling the external lines by
(1)\,-\,(4), the internal propagators by (5) and (6) and the weak
vertex by (7). The mass singularities at NLO arise from diagrams
in which a gluon is attached to two of the circles in
fig.~\ref{fig:1loop8} (including diagrams in which both ends of
the gluon are attached to the same circle). We denote by $(ij)$
the diagram obtained by adding a gluon between the circles $(i)$
and $(j)$ (see, for example, fig.~\ref{fig:diags1} where diagrams
(34), (23) and (13) are drawn explicitly). We evaluate the terms
containing the mass singularities and/or large logarithms which
are of the form $\log(M_B^2/q_2.k)$.

We present the results from an explicit evaluation of each diagram
in Appendix~\ref{app:explicit}, from which the cancellation of
mass singularities from the hard-scattering kernel is apparent. In
the calculation presented in the appendix we do not make any a
priori assumptions about the regions of phase space which lead to
mass singularities or large logarithms. In this section we
demonstrate the cancellation of mass singularities using a
heuristic argument, based on power counting, which shows that the
mass singularities arise from regions in which the gluon is soft
or collinear with $q_2$ or $k$ (including contributions from the
\textit{soft-collinear} region of phase-space~\cite{SCmodes}). In
Appendix~\ref{app:general} we argue further that such a
cancellation holds for all the weak operators in the effective
Hamiltonian of eq.~(\ref{eq:heff}).

We emphasize that not all the one-loop corrections are given by
the set of diagrams $\{(ij)\}$, and for illustration in
figure~\ref{fig:diagramD} we draw an additional Feynman diagram
(called $D$). However, all the mass singularities are given by
diagrams $(ij)$ and only these diagrams and diagram $D$ give large
logarithms. The remaining diagrams, which may contribute either to
$T^I$ or $T^{II}$, are only sensitive to physics at scales of
$O(M_B)$.

We implement a similar notation for diagrams which give the
one-loop corrections to the distribution amplitudes, and which
therefore contribute to $\Phi^{b\bar{q}^\prime\,(1)} \otimes
T^{(0)} \otimes \Phi^{q\bar{q}^\prime\,(0)}$ and
$\Phi^{b\bar{q}^\prime\,(0)} \otimes T^{(0)} \otimes
\Phi^{q\bar{q}^\prime\,(1)}$ (see eq.\,(\ref{eq:t1def})\,). By
$(13)^\prime$, $(11)^\prime$ and $(33)^\prime$ we denote the
contribution to $\Phi^{b\bar{q}^\prime\,(1)} \otimes T^{(0)}
\otimes \Phi^{q\bar{q}^\prime\,(0)}$ in which the gluon in the
one-loop correction to the distribution amplitude of the initial
$b\bar{q}^\prime$ state ($\Phi^{b\bar{q}^\prime\,(1)}$) is
attached to lines 1 and 3 as indicated (where 1 denotes the
$b$-quark and $3$ the spectator antiquark). Similarly by
$(24)^\prime$, $(22)^\prime$ and $(44)^\prime$ we denote the
contribution to $\Phi^{b\bar{q}^\prime\,(0)} \otimes T^{(0)}
\otimes \Phi^{q\bar{q}^\prime\,(1)}$ in which the gluon in the
one-loop correction to the distribution amplitude of the final
$q\bar{q}^\prime$ state ($\Phi^{q\bar{q}^\prime\,(1)}$) is
attached to lines 2 and 4 as indicated (where 2 and 4 denote the
quark and spectator antiquark respectively). There are also
corrections to the distribution amplitudes in which one end of the
gluon is attached to the path-ordered exponential. By
$(1B)^\prime$ ($(3B)^\prime$) we denote the contribution to
$\Phi^{b\bar{q}^\prime\,(1)} \otimes T^{(0)} \otimes
\Phi^{q\bar{q}^\prime\,(0)}$ in which one end of the gluon in the
one-loop correction to the distribution amplitude of the initial
$b\bar{q}^\prime$ state is attached to line 1 (line 3) and the
other to the path-ordered exponential. Similarly by $(2V)^\prime$
($(4V)^\prime$) we denote the contribution to
$\Phi^{b\bar{q}^\prime\,(0)} \otimes T^{(0)} \otimes
\Phi^{q\bar{q}^\prime\,(1)}$ in which one end of the gluon in the
one-loop correction to the distribution amplitude of the final
$q\bar{q}^\prime$ state is attached to line 2 (line 4) and the
other to the path-ordered exponential.

We now present our heuristic diagrammatic argument for the
cancellation of the mass singularities. In
Appendix~\ref{app:general} we give a formal argument explaining
the reasons for this cancellation based on the collinear Ward
identity.

\subsection{Cancellation of Soft Divergences}\label{subsec:softO8}

We start by considering soft divergences, which arise when the
loop momentum is small, with all components vanishing in a similar
way. Such divergences arise in diagrams with a gluon attached to
two external lines. As explained below, some of these divergences
are absorbed into the mesons' distribution amplitudes in a
straightforward manner. The cancellation of the remaining infrared
divergences occurs as expected from arguments based on
\textit{colour transparency}~\cite{colour_transparency}. Soft
gluons only couple to the compact, colour-singlet vector state
through a dipole interaction, which is non-singular. In terms of
Feynman diagrams there is a cancellation of infrared divergences
between graphs in which the soft gluons couple to different
constituents of the vector meson.

To illustrate this point consider the graph~(34), drawn explicitly
in fig.\,\ref{fig:diags1}, which is singular in the soft region,
$l^\mu\to 0$. The contribution from this region is
\begin{eqnarray}
(34)&=&-ig^2\left(C_F-\frac{C_A}{2}\right)\,{\cal
A}_8^{(0)}\int\frac{d^4l}{(2\pi)^4}
\frac{4(k-l)\cdot xq_2}{l^2(l-k)^2(xq_2-l)^2}\nonumber\\
&\simeq& 2ig^2\left(C_F-\frac{C_A}{2}\right)\,{\cal
A}_8^{(0)}\int\frac{d^4l}{(2\pi)^4} \frac{(k-l)\cdot
q_2}{l^2(l-k)^2\,(q_2.l)}\,.
\end{eqnarray}

In the soft region the contribution from diagram (23) (also drawn
in fig.\,\ref{fig:diags1}) is
\begin{eqnarray}
(23)&=&ig^2\left(C_F-\frac{C_A}{2}\right)\,{\cal
A}_8^{(0)}\int\frac{d^4l}{(2\pi)^4}
\frac{4(k-l)\cdot \bar{x}q_2}{l^2(l-k)^2(\bar{x}q_2-l)^2}\nonumber\\
&\simeq&-2ig^2\left(C_F-\frac{C_A}{2}\right)\,{\cal
A}_8^{(0)}\int\frac{d^4l}{(2\pi)^4} \frac{(k-l)\cdot
q_2}{l^2(l-k)^2\,(q_2.l)}=-(34).
\end{eqnarray}
The soft divergences in diagrams (12) and (14) cancel in an
analogous way. In addition there are soft divergences in the
self-energy diagrams $(nn)$ ($n$=1\,-\,4), but these are trivially
absorbed into the $B$ and vector-meson light-cone wave functions
(diagrams $(nn)^\prime$, with $n$=1\,-\,4).

\begin{center}
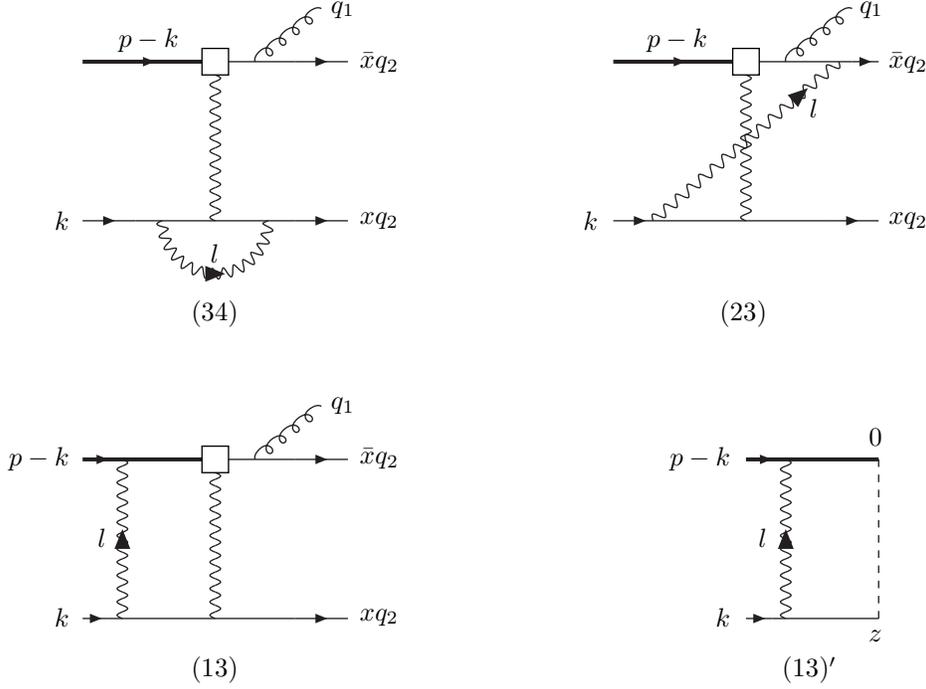
\begin{figure}[t]
\begin{picture}(300,250)(-70,-180)
\ArrowLine(0,0)(20,0) \Line(20,0)(80,0)\ArrowLine(80,0)(100,0)
\SetWidth{1.5}\ArrowLine(49.5,-20)(50.5,-20)\Line(0,60)(50,60)\SetWidth{0.5}
\Line(50,60)(80,60) \ArrowLine(80,60)(100,60)
\ArrowLine(24,60)(26,60) \Text(50,-16)[b]{\small$l$}
\Photon(50,60)(50,0){2}{12} \Gluon(65,60)(90,80){2}{4}
\Text(-5,0)[r]{\small{$k$}}\Text(105,0)[l]{\small{$xq_2$}}
\Text(25,64)[b]{\small{$p-k$}}\Text(105,60)[l]{\small{$\bar{x}q_2$}}
\Text(94,80)[l]{\small{$q_1$}}\GBox(45,55)(55,65){1}
\PhotonArc(50,0)(20,180,360){2}{14} \Text(50,-30)[t]{\small{(34)}}
\ArrowLine(200,0)(220,0)
\Line(220,0)(280,0)\ArrowLine(280,0)(300,0)
\SetWidth{1.5}\Line(200,60)(250,60)\ArrowLine(270,47.14)(271,48)
\SetWidth{0.5} \Line(250,60)(290,60) \ArrowLine(290,60)(300,60)
\ArrowLine(224,60)(226,60) \Text(275,46)[lt]{\small$l$}
\Photon(250,60)(250,0){2}{12} \Gluon(265,60)(290,80){2}{4}
\Text(195,0)[r]{\small{$k$}}\Text(305,0)[l]{\small{$xq_2$}}
\Text(225,64)[b]{\small{$p-k$}}\Text(305,60)[l]{\small{$\bar{x}q_2$}}
\Text(294,80)[l]{\small{$q_1$}}\GBox(245,55)(255,65){1}
\Photon(215,0)(285,60){2}{16} \Text(250,-30)[t]{\small{(23)}}
\ArrowLine(0,-150)(10,-150)
\Line(10,-150)(80,-150)\ArrowLine(80,-150)(100,-150)
\SetWidth{1.5}\ArrowLine(15,-120.5)(15,-119.5)
\Line(0,-90)(50,-90)\SetWidth{0.5}\ArrowLine(7,-90)(8,-90)
\Line(50,-90)(80,-90) \ArrowLine(80,-90)(100,-90)
\Text(9,-120)[r]{\small$l$} \Photon(50,-90)(50,-150){2}{12}
\Gluon(65,-90)(90,-70){2}{4}
\Text(-5,-150)[r]{\small{$k$}}\Text(105,-150)[l]{\small{$xq_2$}}
\Text(-5,-90)[r]{\small{$p-k$}}\Text(105,-90)[l]{\small{$\bar{x}q_2$}}
\Text(94,-70)[l]{\small{$q_1$}}\GBox(45,-95)(55,-85){1}
\Photon(15,-90)(15,-150){2}{12} \Text(50,-165)[t]{\small{(13)}}
\ArrowLine(250,-150)(260,-150) \Line(260,-150)(300,-150)
\SetWidth{1.5}\Line(250,-90)(300,-90)\ArrowLine(265,-120.5)(265,-119.5)
\SetWidth{0.5} \ArrowLine(257,-90)(258,-90)
\Text(259,-120)[r]{\small$l$}
\DashLine(300,-90)(300,-150){3}\Text(245,-150)[r]{\small{$k$}}
\Text(245,-90)[r]{\small{$p-k$}}\Photon(265,-90)(265,-150){2}{12}
\Text(275,-165)[t]{\small{(13)$^\prime$}}\Text(300,-85)[b]{\small{0}}
\Text(300,-155)[t]{\small{$z$}}
\end{picture}
\caption{The one-loop graphs (34), (23) and (13) contributing to
the decay amplitude and a diagram which contributes to the
distribution amplitude of the initial state
$\Phi^{b\bar{q}^\prime\,(1)}$\,. The dashed line represents the
path-ordered exponential and $z^2=0$.}\label{fig:diags1}
\end{figure}
\end{center}\vspace{-40pt}

Before proceeding to a discussion of collinear divergences, we
briefly comment on diagram (13) drawn in fig.\,\ref{fig:diags1}.
By power counting one can readily demonstrate that this diagram
has a leading-twist contribution from the soft region in which the
components of $l$ are $O(\Lambda_{{\scriptsize\textrm{QCD}}})$.
This contribution is also absorbed into the corresponding
distribution amplitude; specifically the contribution from diagram
(13) is cancelled by the subtraction of the corresponding term in
$\Phi^{b\bar{q}^\prime\,(1)}\otimes T^{(0)}\otimes
\Phi^{q\bar{q}^\prime\,(0)}$ (i.e. by the subtraction of diagram
$(13)^\prime$ in which the diagram contributing to
$\Phi^{b\bar{q}^\prime\,(1)}$ is also drawn in
fig.\,\ref{fig:diags1}). The reason for such cancellations has
been explained in detail in ref.~\cite{dgs_glnu}. Diagram
$(13)^\prime$ differs from (13) in that the propagator of the
gluon attached to the weak vertex is replaced by its eikonal
approximation ($(xq_2-k+l)^2\to -2xq_2\cdot(k-l)$). Since the
leading-twist contribution in (13) comes from the soft region, we
can approximate this propagator by the eikonal form in this
diagram. Thus the two contributions are equal and therefore cancel
in the evaluation of the hard-scattering amplitude. Diagrams (24)
and (24)$^\prime$ do not have such leading-twist contributions
from the soft region.

\subsection{Cancellation of Collinear
Divergences}\label{subsec:collinearO8}

We now consider collinear divergences, present when a gluon is
attached to an external light-quark line. Such singularities can
occur when the gluon is parallel to the momentum of the vector
meson $q_2$ or to the momentum of the spectator antiquark $k$ (in
this latter case, since $k=O(\lqcd)$, one does not call the
corresponding gluons \textit{collinear} in the standard SCET
nomenclature). We distinguish between these two cases, beginning
our discussion with the former.

\subsubsection{Divergences from the Region Collinear to
\boldmath{$q_2$}}\label{subsubsec:collinearO8q2}

In this section we consider the region of phase space in which the
loop momentum is parallel to $q_2$. There are corresponding mass
singularities in diagrams with a gluon attached to one of the
final-state external lines, 2 or 4. Before evaluating each diagram
in turn, it is convenient to consider the distribution amplitude
of the final state, in this case the state with a light quark and
a light anti-quark with momenta $\bar{x}q_2$ and $xq_2$
respectively:
\begin{equation}
\Phi^{q\bar{q}^\prime}_{\alpha\beta}(u)=q_{2-}\int\,\frac{dz_+}{2\pi}
\,e^{-i(uq_2\cdot x +\bar{u}q_2\cdot y)}\
\langle\bar{q}^\prime(xq_2)\,q(\bar{x}q_2)\,|\bar{q}^\prime_\beta(y)
[y,x]q_\alpha(x)\,|\,0\rangle\,,
\end{equation}
where $z=x-y$, $\alpha$ and $\beta$ are spinor labels and $[y,x]$
represents the path-ordered exponential between the two points $y$
and $x$.

\begin{center}
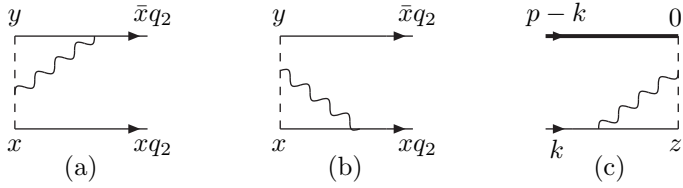
\begin{figure}[t]
\begin{picture}(250,65)(-100,-20)
\Line(0,0)(40,0)\ArrowLine(40,0)(50,0)
\Line(0,35)(40,35)\ArrowLine(40,35)(50,35) \DashLine(0,0)(0,35){3}
\Photon(0,13)(30,35){2}{4} \Text(0,-4)[t]{\small{$x$}}
\Text(0,39)[b]{\small{$y$}} \Text(45,-4)[tl]{\small{$xq_2$}}
\Text(45,39)[bl]{\small{$\bar{x}q_2$}}\Text(25,-10)[t]{\small(a)}
\Line(100,0)(140,0)\ArrowLine(140,0)(150,0)
\Line(100,35)(140,35)\ArrowLine(140,35)(150,35)
\DashLine(100,0)(100,35){3} \Photon(100,22)(130,0){2}{4}
\Text(100,-4)[t]{\small{$x$}} \Text(100,39)[b]{\small{$y$}}
\Text(145,-4)[tl]{\small{$xq_2$}}
\Text(145,39)[bl]{\small{$\bar{x}q_2$}}\Text(125,-10)[t]{\small(b)}
\ArrowLine(200,0)(210,0)\Line(210,0)(250,0)\ArrowLine(204.5,35)(205.5,35)
\SetWidth{1.5} \Line(200,35)(250,35)\SetWidth{0.5}
\DashLine(250,0)(250,35){3} \Photon(250,22)(220,0){2}{4}
\Text(250,-4)[t]{\small{$z$}} \Text(250,39)[b]{\small{$0$}}
\Text(205,-4)[t]{\small{$k$}} \Text(205,39)[b]{\small{$p-k$}}
\Text(225,-10)[t]{\small(c)}
\end{picture}
\caption{Two one-loop diagrams contributing to the distribution
amplitude of the $q\bar{q}^\prime$ final state, (a) and (b), and a
diagram contributing to the distribution amplitude of the
$b\bar{q}^\prime$ final state. The dashed line represents the
path-ordered exponential and $z^2=(x-y)^2=0$.\label{fig:da1loop}}
\end{figure}
\end{center}
\vspace{-0.6in}We have seen above that for the free theory (see
eq.~(\ref{eq:phim0})\,)
\begin{equation}
\Phi^{q\bar{q}^\prime\,(0)}_{\alpha\beta}(u)=\delta(u-x)\
\bar{u}_\beta(\bar{x}q_2)\, v_\alpha(xq_2)\,,
\end{equation}
where $\bar{u}_\beta$ and $v_\alpha$ are free Dirac spinors. Now
consider the contributions to the distribution amplitude from the
two one-loop diagrams in fig.~\ref{fig:da1loop}(a) and (b), in
which a gluon is attached at one end to the path-ordered
exponential~\footnote{The remaining one-loop diagrams contributing
to the distribution amplitude will be discussed later.}. In the
collinear region in which $l$, the momentum of the gluon, is
parallel to $q_2$ ($l\simeq \xi q_2$, with the remaining
components of $l$ vanishing as $l_+\sim\lambda^2/l_-$ and
$l_\perp\sim\lambda$ as $\lambda\to 0$) their contributions can be
written as:
\begin{eqnarray}\label{eq:phi1a}
\Phi^{q\bar{q}^\prime\,(1)}_{(a)\,\alpha\beta}(u)&=&\bar{u}_\beta(\bar{x}q_2)\,
v_\alpha(xq_2)\ \times\nonumber\\
&&\hspace{-0.1in}(2ig^2\,C_F)\,\int\,
\frac{d^4l}{(2\pi)^4}\frac{\bar{x}-\xi}{\xi}\,\frac{1}{l^2}\frac{1}{(\bar{x}q_2-l)^2}\,
\left\{\delta(u-x)-\delta(u-x-\xi)\right\}\,,\\
\Phi^{q\bar{q}^\prime\,(1)}_{(b)\,\alpha\beta}(u)&=&\bar{u}_\beta(\bar{x}q_2)\,
v_\alpha(xq_2)\ \times\nonumber\\
&&\hspace{-0.1in}(2ig^2\,C_F)\,\int\,
\frac{d^4l}{(2\pi)^4}\frac{x-\xi}{\xi}\,\frac{1}{l^2}\frac{1}{(xq_2-l)^2}\,
\left\{\delta(u-x)-\delta(u-x+\xi)\right\}\,,
\label{eq:phi1b}\end{eqnarray} where the subscripts $(a)$ and
$(b)$ refer to the corresponding diagrams in
fig.~\ref{fig:da1loop}. We will use the representations in
eqs.~(\ref{eq:phi1a}) and (\ref{eq:phi1b}) in our discussion
below. For the other diagrams contributing to the wave function of
the final state at one-loop order it will be very straightforward
to see how the mass singularities are precisely those needed to
absorb the corresponding ones from the amplitude.

In order to illustrate the cancellation of collinear divergences
coming from the region of phase space in which the momentum of
gluon(s) in the loop is(are) parallel to $q_2$ consider diagram
(12). Since we are considering the collinear divergence we set
$l\simeq\xi q_2$, with $\xi$ finite and the remaining components
of $l$ vanish as $l_+\sim\lambda^2/l_-$ and $l_\perp\sim\lambda$
in the singular region $\lambda\to 0$. Keeping only the
leading-twist terms one readily finds
\begin{equation}\label{eq:collq212}
(12)=2ig^2\left(C_F-\frac{C_A}{2}\right){\cal A}_8^{(0)}
\int\frac{d^4l}{(2\pi)^4}\,\frac{\bar{x}}{\xi}\,\frac{1}{l^2}\,
\frac{1}{(\bar{x}q_2-l)^2}\,.
\end{equation}
The integral over $l_+$ and $l_\perp$ gives a collinear
divergence. These collinear divergences also cancel, but in a
different way to the soft divergences. In this case it is
straightforward to verify that in the region where $l$ is
collinear to $q_2$,
\begin{equation}\label{eq:collq223}
(23)=-(12)\,,
\end{equation}
so that the corresponding collinear divergences cancel between
diagrams (12) and (23). As we will see when we consider the
cancellation of collinear divergences for the operator ${\cal
Q}_1$ below, such a cancellation between pairs of diagrams is not
typical. What is required however, is that apart from the
propagators which are explicitly exhibited in
eq.~(\ref{eq:collq212}) ($1/l^2$ and $1/(\bar{x}q_2-l)^2$) one can
replace $l$ by $\xi q_2$ everywhere.

There are similar cancellations between other pairs of
diagrams~\footnote{The mass singularities considered in this
section appear in diagrams in which a gluon in a loop is attached
to one of the final-state external lines 2 or 4. Below however, we
give singular expressions for diagrams (36) and (37). These
diagrams could equally well have been denoted by (46) and (47) and
hence are indeed singular.}
\begin{eqnarray}
(14)=-(34)&=&-2ig^2\left(C_F-\frac{C_A}{2}\right){\cal A}_8^{(0)}
\int\frac{d^4l}{(2\pi)^4}\,\frac{x-\xi}{\xi}\,\frac{1}{l^2}\,
\frac{1}{(xq_2-l)^2}\,;\label{eq:collq214}\\
(26)=-(27)&=&ig^2C_A {\cal A}_8^{(0)}\ \frac{\bar{x}}{x}
\int\frac{d^4l}{(2\pi)^4}\,\frac{1}{l^2}\,
\frac{1}{(\bar{x}q_2-l)^2}\,;\\
(36)=-(37)&=&-ig^2C_A {\cal A}_8^{(0)}
\int\frac{d^4l}{(2\pi)^4}\,\frac{1}{l^2}\, \frac{1}{(xq_2-l)^2}\,.
\end{eqnarray}

This leaves us with the collinear divergences from diagrams (25)
and (45) in which one end of the additional gluon is attached to
the internal quark propagator 5. These are precisely the terms
which are absorbed into the wave function of the final state (in
this case the $q\bar{q}^\prime$ state), i.e. the divergences in
the contributions from diagrams (25) and (45) to ${\cal
A}_8^{(1)}$ are cancelled by the subtraction of
$\Phi^{b\bar{q}^\prime\,(0)} \otimes T_8^{(0)} \otimes
\Phi^{q\bar{q}^\prime\,(1)}$ in eq.~(\ref{eq:t1def}). More
specifically they are cancelled by the contributions to the wave
function from the diagrams in fig.~\ref{fig:da1loop}(a) and (b),
which we denote by $(2V)^\prime$ and $(4V)^\prime$ respectively.
Using the expressions in eqs.~(\ref{eq:phi1a}) and
(\ref{eq:phi1b}), it is straightforward to demonstrate that in
this collinear region
\begin{eqnarray}
(25)&=&(2V)^\prime=-2ig^2C_F\,{\cal
A}_8^{(0)}\int\frac{d^4l}{(2\pi)^4}\,\frac{1}{l^2}\,
\frac{1}{(\bar{x}q_2-l)^2}\,;\\
(45)&=&(4V)^\prime=2ig^2C_F\,{\cal
A}_8^{(0)}\int\frac{d^4l}{(2\pi)^4}\,
\frac{x-\xi}{\bar{x}+\xi}\,\frac{1}{l^2}\, \frac{1}{(xq_2-l)^2}\,.
\end{eqnarray}

Finally we discuss the contribution of diagram (24). For the
processes considered in this paper, with a transversely polarized
vector meson in the final state, the leading-twist contribution
vanishes. We therefore do not discuss this further, other than to
note that for a generic final state this contribution is cancelled
by the corresponding one in $\Phi^{b\bar{q}^\prime\,(0)} \otimes
T_8^{(0)} \otimes \Phi^{q\bar{q}^\prime\,(1)}$.

Thus we find that, at one-loop order, there are no singular
contributions to $T_8^{(1)}$ from the region in which the momentum
of the gluon $l$ is parallel to $q_2$, and that the cancellation
of the singularities has a very simple structure. The divergences
either cancel between pairs of diagrams contributing to ${\cal
A}_8^{(1)}$, or between diagrams contributing to ${\cal
A}_8^{(1)}$ and to $\Phi^{b\bar{q}^\prime\,(0)} \otimes T_8^{(0)}
\otimes \Phi^{q\bar{q}^\prime\,(1)}$.

\subsubsection{Divergences from the Region Collinear to
\boldmath{$k$}}\label{subsubsec:o8collineark}

For the $B$-meson initial state, the momentum of the spectator
antiquark, $k$, is generically soft, with all components of
$O(\lqcd)$. In our calculations we take $k$ to be light-like so
that long-distance effects are manifested by the presence of mass
singularities, specifically collinear divergences from the region
of phase space where the momenta of internal partons are collinear
with $k$ (as well as the soft divergences discussed above). We now
consider these collinear divergences. The diagrams which can have
such a divergence contain a gluon line attached to the external
line 3. Writing $l\simeq\sigma k$ with $l_-$ and $l_\perp$
vanishing with $\lambda$ as $\lambda^2/l_+$ and $\lambda$
respectively, we readily find the following three leading-twist
contributions (diagram (37)=0 and (35) gives a higher-twist
contribution):
\begin{eqnarray}
(23)&=&-2ig^2\left(C_F-\frac{C_A}{2}\right){\cal
A}_8^{(0)}\int\frac{d^4l}{(2\pi)^4}\,
\frac{1}{\sigma}\,\frac{1}{l^2\,(k-l)^2}\,,\label{eq:collk23}\\
(34)&=&\ \ \, 2ig^2\left(C_F-\frac{C_A}{2}\right){\cal
A}_8^{(0)}\int\frac{d^4l}{(2\pi)^4}\,
\frac{1-\sigma}{\sigma}\,\frac{1}{l^2\,(k-l)^2}\,,\label{eq:collk34}\\
(36)&=&-ig^2C_A\,{\cal A}_8^{(0)}\int\frac{d^4l}{(2\pi)^4}\,
\frac{1}{l^2\,(k-l)^2}\,.
\end{eqnarray}
Summing these contributions we obtain
\begin{equation}\label{eq:collk3tot}
(23)+(34)+(36)=-2ig^2C_F\,{\cal
A}_8^{(0)}\int\frac{d^4l}{(2\pi)^4}\,
\frac{1}{l^2\,(k-l)^2}=(3B)^\prime,
\end{equation}
where $(3B)^\prime$ is the corresponding one-loop contribution to
$\Phi^{b\bar{q}^\prime\,(1)} \otimes T_8^{(0)} \otimes
\Phi^{q\bar{q}^\prime\,(0)}$, i.e. the one from the contribution
of the graph in fig.~\ref{fig:da1loop}(c) to
$\Phi^{b\bar{q}^\prime\,(1)}$. Thus there are no collinear
singularities in $T_8^{(1)}$ arising from the region in which $l$
is parallel to $k$.

\subsection{Double Logarithms}\label{subsec:doublelogs}

In sections \ref{subsec:softO8} and \ref{subsec:collinearO8} we
have analyzed the regions of phase space in which the momentum of
a gluon in the loop was either soft or collinear with $q_2$ or
$k$, and deduced that the cancellation of the corresponding
divergences is straightforward. We now briefly analyze the region
in which the gluons are both soft and collinear, which leads to
double logarithms in an infrared cut-off for some of the diagrams.

The double logarithms arise in diagrams (12) and (14) as $\xi\to
0$ (see equations (\ref{eq:collq212}) and (\ref{eq:collq214})) and
in diagrams (23) and (34) as either $\xi\to 0$
(equations~(\ref{eq:collq212}), (\ref{eq:collq223}) and
(\ref{eq:collq214})\,) or $\sigma\to 0$
(equations~(\ref{eq:collk23}) and (\ref{eq:collk34})). Explicit
evaluation of the diagrams, described in Appendix
\ref{app:explicit}, confirms that only these four diagrams contain
double logarithms.

From the expressions in the previous subsections we can also
deduce how the double logarithms cancel. We know from
section~\ref{subsec:softO8} that the soft divergences cancel in
the sum of diagrams (23)+(34), and from equations
(\ref{eq:collk23}) and (\ref{eq:collk34}) we note that the term
which diverges as $1/\sigma$ as $\sigma\to 0$ cancels in the
integrand of (23)+(34). Thus the double logarithms in diagrams
(23) and (34) cancel, leaving only a collinear divergence from the
region in which $l$ is parallel to $q_2$. This is confirmed by
explicit calculations (see eqs.~(\ref{eq:app23ab}) and
(\ref{eq:app34ab}) in Appendix \ref{app:explicit}). Similarly,
since the soft divergences cancel in diagrams (12) and (14), in
the sum of these two diagrams we are only left with a single
logarithm corresponding to a collinear divergence from the region
in which $l$ is parallel to $q_2$.

When presenting such heuristic arguments, one has to consider the
possibility that, after the cancellation of the double logarithms,
a subtlety in the way that the integrals are regulated may leave a
single logarithm which had not been considered. The explicit
calculations in appendix \ref{app:explicit} clearly remove such
worries, but it is also relatively straightforward to verify this
without the full computations. For example one may evaluate either
the $l_-$ or $l_+$ integration by contours, and readily verify the
above pattern of cancellations.

\section{Factorisation at One-Loop Order for the Four-Quark
Operator \boldmath{${\cal Q}_1$}} \label{sec:fact1lpo1}

In this section we repeat the heuristic arguments with ${\cal
Q}_1$ as the weak operator. For the soft divergences which are
present in diagrams in which the additional gluon is attached to
two external lines, the cancellation occurs in the standard way,
exactly as described in section~\ref{subsec:softO8} and we don't
discuss them any further. For the collinear divergences, we
demonstrate below that they also cancel, but the pattern of
cancellation is different in detail from that in
section~\ref{sec:fact1lpo8}. Moreover the cancellation occurs even
if we keep higher-twist contributions such as $m_c^2/m_b^2$ or
$2q_2\cdot k/m_b^2$ in the quark-loop. This is consistent with the
the arguments based on collinear Ward identities presented in
appendix~\ref{app:general}.

\begin{figure}[t]
\begin{center}
\begin{picture}(230,150)(-15,-5)
\SetWidth{1.5}\Line(0,100)(80,100)\SetWidth{0.8}
\ArrowLine(9,100)(10,100)\SetWidth{0.5}
\Text(-5,100)[r]{\small{$p$\,--\,$k$}}
\GCirc(40,100){7}{1}\Text(40,100)[]{1}
\Line(80,100)(160,100)\ArrowLine(150,100)(151,100)
\GCirc(120,100){7}{1}\Text(120,100)[]{2}
\Text(165,100)[l]{\small{$\bar{x}q_2$}}
\Line(0,0)(160,0)\ArrowLine(9,0)(10,0)\ArrowLine(150,0)(151,0)
\Text(165,0)[l]{\small{$xq_2$}} \Text(-5,0)[r]{\small{$k$}}
\GCirc(40,0){7}{1}\Text(40,0)[]{3}
\GCirc(120,0){7}{1}\Text(120,0)[]{4}
\GCirc(80,100){2}{0}\GCirc(80,95){2}{0} \Oval(80,70)(25,15)(0)
\Gluon(95,70)(130,70){2.5}{4}\Text(135,70)[l]{\small{$q_1$}}
\GCirc(80,70){7}{1}\Text(80,70)[]{7} \Photon(80,45)(80,0){3}{6.5}
\GCirc(80,22.5){7}{1}\Text(80,22.5)[]{6}

\end{picture}
\caption{Notation for the next-to-leading order diagrams
contributing to the $\bvg$ decay from ${\cal Q}_1$ (the diagram
with the photon emitted from the other propagator in the quark
loop is included implicitly). The extra gluon can be attached to
any pair of circles (which might be identical). $(n7)$, for
$n=1$\,--\,6, represents the four diagrams for which the external
photon and internal gluon are attached to either the quark or
antiquark in the loop.} \label{fig:1loop1}
\end{center}
\end{figure}
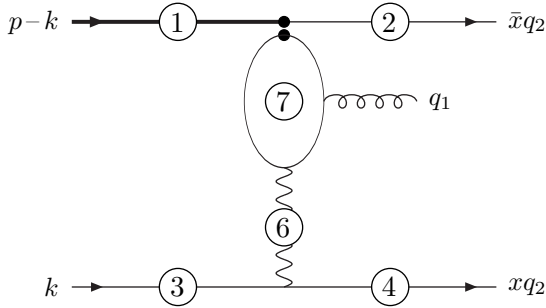
The notation for the next-to-leading order diagrams is similar to
that in our discussion of the contributions from ${\cal Q}_8$. We
denote by $(ij)$ the diagrams in which the external photon is
emitted from either of the two propagators in the quark loop and
the internal gluon is attached to circles $i$ and $j$. By $(n7)$
we mean the sum of the diagrams with one end of the gluon attached
to either the quark or antiquark in the loop.

\subsection{Divergences from the Region Collinear to
\boldmath{$q_2$}}

\subsubsection{Neglecting terms of \boldmath{$O(m_c^2/m_b^2)$}}
\label{subsubsec:mceq01} We now repeat the steps of section
\ref{subsubsec:collinearO8q2} for the operator ${\cal Q}_1$ with
the photon emitted from the internal loop, beginning with the
simpler case in which we neglect $m_c$. As before we start with
diagram (12) (see fig.~\ref{fig:o11223}) for which the expression
in this collinear region is now
\begin{equation}\label{eq:o1collq212}
(12)=2ig^2\left(C_F-\frac{C_A}{2}\right){\cal A}^{(0)}_1
\int\frac{d^4l}{(2\pi)^4}\,\frac{\bar{x}-\xi}{\xi}\,\frac{1}{l^2}\,
\frac{1}{(\bar{x}q_2-l)^2}\,,
\end{equation}
where ${\cal A}_1^{(0)}$ is the amplitude at tree level. This is
no longer cancelled by the corresponding contribution from diagram
(23) (see fig.~\ref{fig:o11223}) which is now given by
\begin{equation}\label{eq:o1collq223}
(23)=-2ig^2\left(C_F-\frac{C_A}{2}\right){\cal A}^{(0)}_1
\int\frac{d^4l}{(2\pi)^4}\,\frac{\bar{x}-\xi}{\xi}\,\frac{x}{x+\xi}\,\frac{1}{l^2}\,
\frac{1}{(\bar{x}q_2-l)^2}\,.
\end{equation}
Using eq.~(\ref{eq:phi1a}), we can see that the contribution
$(2V)^\prime$ to $\Phi^{b\bar{q}^\prime\,(0)} \otimes T_8^{(0)}
\otimes \Phi^{q\bar{q}^\prime\,(1)}$ cancels the abelian component
of these two diagrams:
\begin{equation}
(2V)^\prime = 2ig^2\,C_F\,{\cal A}^{(0)}_1
\int\frac{d^4l}{(2\pi)^4}\,\frac{\bar{x}-\xi}{x+\xi}\,\frac{1}{l^2}\,
\frac{1}{(\bar{x}q_2-l)^2}\,.
\end{equation}
Indeed the abelian component of diagram (12) is cancelled by the
contribution to $(2V)^\prime$ coming from the first delta function
in eq.~(\ref{eq:phi1a}) and similarly the collinear divergence in
diagram (23) is cancelled by the contribution from the second
delta function. This simple observation will be important when we
generalise the discussion to include corrections of
$O(m_c^2/m_b^2)$. Thus we have
\begin{equation}\label{eq:nonabelian2}
(12)+(23)-(2V)^\prime=-ig^2\,C_A\,{\cal A}^{(0)}_1
\int\frac{d^4l}{(2\pi)^4}\,\frac{\bar{x}-\xi}{x+\xi}\,\frac{1}{l^2}\,
\frac{1}{(\bar{x}q_2-l)^2}\,.
\end{equation}
In the case where we neglect the mass of the charm quark, the
non-abelian contribution in eq.~(\ref{eq:nonabelian2}) is
cancelled by diagram (26). There are no collinear divergences from
the sum of diagrams (27).

Similarly, the expressions for diagrams (34) (which is unchanged
from that in eq.~(\ref{eq:collq214})), (14) and $(4V)^\prime$ (see
eq.~(\ref{eq:phi1b})\,) are now
\begin{eqnarray}
(34)&=&2ig^2\left(C_F-\frac{C_A}{2}\right){\cal A}^{(0)}_1
\int\frac{d^4l}{(2\pi)^4}\,\frac{x-\xi}{\xi}\,\frac{1}{l^2}\,
\frac{1}{(xq_2-l)^2}\,,\\
(14)&=&-2ig^2\left(C_F-\frac{C_A}{2}\right){\cal A}^{(0)}_1
\int\frac{d^4l}{(2\pi)^4}\,\frac{x}{\xi}\,\frac{1}{l^2}\,
\frac{1}{(xq_2-l)^2}\quad\textrm{and}\\
(4V)^\prime&=& -2ig^2\,C_F\,{\cal A}^{(0)}_1
\int\frac{d^4l}{(2\pi)^4}\,\frac{1}{l^2}\, \frac{1}{(xq_2-l)^2}\,,
\end{eqnarray}
so that the abelian component cancels and
\begin{equation}\label{eq:nonabelian4}
(14)+(34)-(4V)^\prime=ig^2\,C_A\,{\cal A}^{(0)}_1
\int\frac{d^4l}{(2\pi)^4}\,\frac{1}{l^2}\,
\frac{1}{(\bar{x}q_2-l)^2}\,.
\end{equation}
This contribution is cancelled by diagram (46). If we neglect the
mass of the charm quark there are no collinear divergences from
the sum of the diagrams (47).

Thus the collinear divergences from the region of phase space in
which $l$ is parallel to $q_2$ either cancel or are absorbed into
the vector meson's distribution amplitude.

\subsubsection{Including terms of \boldmath{$O(m_c^2/m_b^2)$}}
\label{subsubsec:mcneq01}
\begin{center}
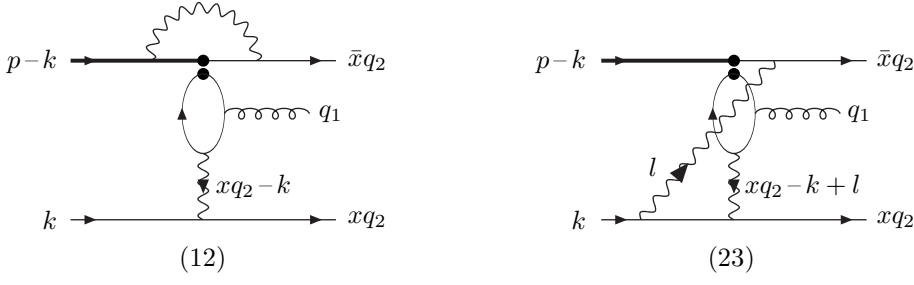
\begin{figure}
\begin{picture}(300,90)(-70,-20)
\Line(0,0)(50,0)\Line(50,0)(100,0) \ArrowLine(7,0)(8,0)
\SetWidth{1.5}\Line(0,60)(50,60)\SetWidth{0.8}
\ArrowLine(50,13)(50,12)\SetWidth{0.5} \Line(50,60)(100,60)
\ArrowLine(92,60)(93,60) \ArrowLine(92,0)(93,0)
\ArrowLine(7,60)(8,60)\GCirc(50,60){2}{0}
\GCirc(50,55){2}{0}\Oval(50,40)(15,8)(0)\Photon(50,25)(50,0){2}{4}
\ArrowLine(42,39.5)(42,40.5)\Gluon(58,40)(90,40){2}{4}
\Text(-5,0)[r]{\small{$k$}}\Text(105,0)[l]{\small{$xq_2$}}
\Text(-5,60)[r]{\small{$p$\,--\,$k$}}\Text(105,60)[l]{\small{$\bar{x}q_2$}}
\Text(94,40)[l]{\small{$q_1$}}\PhotonArc(50,60)(20,0,180){2}{10}
\Text(55,12.5)[l]{\small{$xq_2$\,--\,$k$}}
\Text(50,-10)[t]{\small{(12)}}
\Line(200,0)(250,0)\Line(250,0)(300,0) \ArrowLine(207,0)(208,0)
\SetWidth{1.5}\Line(200,60)(250,60)\ArrowLine(230,18)(231,19.2)\SetWidth{0.8}
\ArrowLine(250,13)(250,12)\SetWidth{0.5} \Line(250,60)(300,60)
\ArrowLine(292,60)(293,60) \ArrowLine(292,0)(293,0)
\ArrowLine(207,60)(208,60)\GCirc(250,60){2}{0}
\GCirc(250,55){2}{0}\Oval(250,40)(15,8)(0)\Photon(250,25)(250,0){2}{4}
\ArrowLine(242,39.5)(242,40.5)\Gluon(258,40)(290,40){2}{4}
\Text(195,0)[r]{\small{$k$}}\Text(305,0)[l]{\small{$xq_2$}}
\Text(195,60)[r]{\small{$p$\,--\,$k$}}\Text(305,60)[l]{\small{$\bar{x}q_2$}}
\Text(294,40)[l]{\small{$q_1$}}\Photon(215,0)(265,60){2}{10}
\Text(255,12.5)[l]{\small{$xq_2$\,--\,$k+l$}}\Text(222,20)[r]{\small{$l$}}
\Text(250,-10)[t]{\small{(23)}}
\end{picture}
\caption{One-loop diagrams (12) and (23) contributing to
corrections to the matrix element of ${\cal Q}_1$. The
corresponding diagrams in which the photon is emitted from the
other propagator in the quark loop are to be considered implicitly
included.}\label{fig:o11223}
\end{figure}
\end{center}\vspace{-40pt}

In this subsection we demonstrate that the collinear divergences
cancel, even if we include the corrections of $O(m_c^2/m_b^2)$.
The additional complication is that the integral over momentum in
the charm-quark loop now depends on whether the outgoing momentum
in the gluon attached to the loop is $xq_2-k$ (as in diagram (12)
in fig.\,\ref{fig:o11223} for example) or $xq_2-k+l$ (as in
diagram (23) in fig.\,\ref{fig:o11223} for example). We will
consider diagrams (27) and (47), in which two gluons are attached
to the charm-quark loop, separately. In order to distinguish
between the two types of contribution we write
\begin{equation}\label{eq:f10}
{\cal A}_1^{(0)}(x)=\frac1x\,I_1(x)\,,
\end{equation}
where the $x$ dependence in $I_1(x)$ comes from the integral over
the loop-momentum in the quark loop and is exhibited explicitly in
eqs.~(\ref{eq:a10})\,--\,(\ref{eq:t3}). If we neglect
$m_c^2/m_b^2$, then $I_1(x)$ is independent of $x$, which leads to
the vanishing of some contributions, such as (27) and (47). Since
now we do not neglect these effects, we must distinguish the two
type of one-loop corrections, corresponding to $I_1(x)$ and
$I_1(x+\xi)$ respectively; in the latter case the factor is in the
integrand of the $l$ integration. We now demonstrate that neither
type of contribution contains collinear divergences.

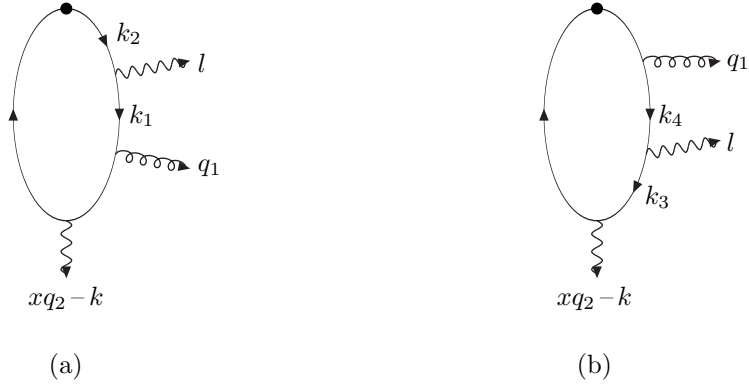
\begin{figure}[t]\begin{center}
\begin{picture}(300,140)(0,-35)
\Oval(50,60)(40,20)(0)\GCirc(50,100){2}{0}\Photon(50,20)(50,0){2}{3}
\ArrowLine(50,1)(50,0)\Text(50,-5)[t]{\small{$xq_2$\,--\,$k$}}
\Gluon(68.54,45)(95,40){2}{4}\Text(100,40)[l]{\small{$q_1$}}
\Photon(68.54,75)(95,80){2}{5}\Text(100,80)[l]{\small{$l$}}
\ArrowLine(94,79.8)(95,80)\ArrowLine(94,40.2)(95,40)
\ArrowLine(30,59.5)(30,60.5)\ArrowLine(70,60.5)(70,59.5)
\Text(74,60)[l]{\small{$k_1$}}
\ArrowLine(64.28,88)(64.75,87)\Text(69,91)[l]{\small{$k_2$}}
\Text(50,-30)[t]{\small{(a)}}
\Oval(250,60)(40,20)(0)\GCirc(250,100){2}{0}\Photon(250,20)(250,0){2}{3}
\ArrowLine(250,1)(250,0)\Text(250,-5)[t]{\small{$xq_2$\,--\,$k$}}
\Photon(268.54,45)(295,50){2}{5}\Text(300,50)[l]{\small{$l$}}
\Gluon(267.32,80)(295,80){2}{4}\Text(300,80)[l]{\small{$q_1$}}
\ArrowLine(294,79.8)(295,80)\ArrowLine(294,49.8)(295,50)
\ArrowLine(230,59.5)(230,60.5)\ArrowLine(270,60.5)(270,59.5)
\Text(274,60)[l]{\small{$k_4$}}
\ArrowLine(264.75,33)(264.28,32)\Text(269,29)[l]{\small{$k_3$}}
\Text(250,-30)[t]{\small{(b)}}
\end{picture}\caption{Two of the attachments of the gluon of momentum $l$
onto the quark loop (see text)\,.\label{fig:loopaux}}
\end{center}
\end{figure}

We start by considering the terms proportional to $I_1(x)$ arising
from diagrams with a gluon attached to the outgoing quark line
with momentum $\bar xq_2$. The corresponding diagrams are (12),
$(2V)^\prime$ and (27). Repeating the steps in
section~\ref{subsubsec:mceq01} we find
\begin{equation}\label{eq:nonabelian2m}
(12)-(2V)^\prime=-ig^2C_A\,\frac{I_1(x)}{x}
\int\frac{d^4l}{(2\pi)^4}\,\frac{\bar{x}-\xi}{\xi}\,\frac{1}{l^2}\,
\frac{1}{(\bar{x}q_2-l)^2}\,.
\end{equation}
The remaining contribution of this type comes from diagrams (27).
The numerator of the light-quark propagator has a factor of
$2(\bar{x}-\xi)q_2^\mu\simeq 2(\bar{x}-\xi)/\xi\, l^\mu$, where
$\mu$ is the Lorentz index of the gluon with momentum $l$.
Consider first the diagram in fig.\,\ref{fig:loopaux}\,(a), for
which $l=k_2-k_1$. Using the Ward identity
\begin{equation}\label{eq:wi0}
\frac{1}{\dirac{k}_1-m}\,\{\dirac{k}_2-\dirac{k}_1\}
\,\frac{1}{\dirac{k}_2-m}=\frac{1}{\dirac{k}_1-m}-
\frac{1}{\dirac{k}_2-m}
\end{equation}
inside the quark loop, we readily see that the integral over the
quark loop can be reduced to ones with only three quark
propagators. The first term on the right-hand side of
eq.\,(\ref{eq:wi0}) gives a contribution proportional to $I_1(x)$.
Similarly consider the insertion of the gluon as in
fig.\,\ref{fig:loopaux}\,(b) and the Ward identity
\begin{equation}\label{eq:wi02}
\frac{1}{\dirac{k}_3-m}\,\{\dirac{k}_4-\dirac{k}_3\}
\,\frac{1}{\dirac{k}_4-m}=\frac{1}{\dirac{k}_3-m}-
\frac{1}{\dirac{k}_4-m}\,.
\end{equation}
The first term on the right-hand side of eq.\,(\ref{eq:wi02})
cancels the second term of eq.\,(\ref{eq:wi0}), and the second
term of eq.\,(\ref{eq:wi02}) gives a contribution proportional to
$I_1(x+\xi)$. These arguments can be generalized to include all
the insertions of the photon and gluon onto the quark loop. The
term proportional to $I_1(x)$ precisely cancels the contribution
from $(12)-(2V)^\prime$ in eq.~(\ref{eq:nonabelian2m}).

The cancellation of the terms with integrands proportional to
$I_1(x+\xi)$ in diagrams with the gluon with momentum $l$ attached
to the external light quark with momentum $\bar{x}q_2$ proceeds in
a similar way. The corresponding contributions are as follows:
\begin{eqnarray*}
(23)-(2V)^\prime&=&ig^2C_A\,
\int\frac{d^4l}{(2\pi)^4}\,\frac{\bar{x}-\xi}{\xi}\,\frac{I_1(x+\xi)}{x+\xi}\,\frac{1}{l^2}\,
\frac{1}{(\bar{x}q_2-l)^2}\,,\\
(26)&=&ig^2C_A\,
\int\frac{d^4l}{(2\pi)^4}\,\frac{\bar{x}-\xi}{x}\,\frac{I_1(x+\xi)}{x+\xi}\,\frac{1}{l^2}\,
\frac{1}{(\bar{x}q_2-l)^2}\,,\\
(27)&=&-ig^2C_A\,
\int\frac{d^4l}{(2\pi)^4}\,\frac{\bar{x}-\xi}{\xi}\,\frac{I_1(x+\xi)}{x}\,\frac{1}{l^2}\,
\frac{1}{(\bar{x}q_2-l)^2}\,,
\end{eqnarray*}
so that
\begin{equation}
(23)-(2V)^\prime+(26)+(27)=0
\end{equation}
and all the mass singularities cancel as required.

The cancellation of the mass singularities from the diagrams in
which the gluon with momentum $l$ is attached to the light
antiquark with momentum $xq_2$ proceeds in a very similar way,
except that now we have factors $I_1(x)$ and $I_1(x-\xi)$. For the
terms with a factor $I_1(x)$ the contributions are:
\begin{eqnarray}
(34)-(4V)^\prime&=&-ig^2C_A\,\frac{I_1(x)}{x}
\int\frac{d^4l}{(2\pi)^4}\,\frac{x-\xi}{\xi}\,\frac{1}{l^2}\,
\frac{1}{(xq_2-l)^2}\,,\\
(46)&=&-ig^2C_A\,\frac{I_1(x)}{x}
\int\frac{d^4l}{(2\pi)^4}\,\frac{1}{l^2}\,
\frac{1}{(xq_2-l)^2}\,,\\
(47)&=&ig^2C_A\,\frac{I_1(x)}{x}
\int\frac{d^4l}{(2\pi)^4}\,\frac{x}{\xi}\,\frac{1}{l^2}\,
\frac{1}{(xq_2-l)^2}\,,\\
\end{eqnarray}
which sum to zero, and for the terms with a factor $I_1(x-\xi)$ in
the integrand we have
\begin{equation}
(14)-(4V)^\prime=-(47)=ig^2C_A\,
\int\frac{d^4l}{(2\pi)^4}\,\frac{I_1(x-\xi)}{\xi}\,\frac{1}{l^2}\,
\frac{1}{(xq_2-l)^2}\,.
\end{equation}

Thus again, all the collinear divergences from the region in which
a gluon's momentum is parallel to $q_2$ cancel.

\subsubsection{Divergences from the Region Collinear to
\boldmath{$k$}}\label{subsubsec:o1collineark}

For the kinematical situation considered in ref.~\cite{BB}, in
which we keep $m_c^2/m_b^2$ but neglect \mbox{$q_2\cdot k/m_b^2$,}
the cancellation of the mass singularities from the region in
which a gluon is collinear to $k$ follows exactly as in the case
of the operator ${\cal Q}_8$ discussed in
section~\ref{subsubsec:o1collineark} (see equations
(\ref{eq:collk23}) - (\ref{eq:collk3tot})\,).

This concludes the demonstration of the cancellation of mass
singularities. However it is also instructive to ask whether they
would have cancelled if we had not neglected $q_2\cdot k/m_b^2$.
We now rewrite the amplitude at lowest order as
\begin{equation}
{\cal A}_1^{(0)}=\frac{1}{xk_+}I_2(x,k_+)\,,
\end{equation}
where $I_2(x,k_+)=k_+I_1(x,k_+)$ and the $k_+$ dependence in
eq.~(\ref{eq:f10}) and subsequent equations was implicit. For all
the discussion above, the $k_+$ dependence in $I_1(x)$ was
contained in a simple overall factor of $1/k_+$. Now, if we keep
terms of $O(q_2\cdot k/m_b^2)$, we have two types of term in the
integrands of the loop integrations, those proportional to
$I_2(x,k_+)$ and those proportional to $I_2(x,(1-\sigma)k_+)$. We
now demonstrate that the corresponding mass singularities cancel
also in this case.

The distribution amplitude of the initial state is defined by
\begin{equation}
\Phi^{b\bar{q}^\prime}_{\alpha\beta}(\tp)=\int dz_-\ e^{i\tp z_-}\
\langle 0|\, \bar{q}^\prime_\beta(z)[z,0]
b_\alpha(0)\,|b(p)\,\bar{q}^\prime(k)\rangle|_{z_+,z_\perp=0},
\end{equation}
so that at tree level (see eq.~(\ref{eq:phih0}))
\begin{equation}
\Phi^{b\bar{q}^\prime\,(0)}_{\alpha\beta}(\tp)=2\pi\delta(k_+-\tp)\,\bar{v}_\beta(k)
u_\alpha(p-k)\,,
\end{equation}
where $v$ and $u$ are the free spinor wave functions of the light
antiquark and the $b$-quark respectively. In the collinear region
which we are considering here, the one-loop correction from the
diagram in fig.~\ref{fig:da1loop}(c) can be written in the form
\begin{equation}
\Phi^{b\bar{q}^\prime\,(1)}_{(c)\,\alpha\beta}(\tp)=2ig^2C_F\,\bar{v}_\beta(k)
u_\alpha(p)\int \frac{d^4l}{(2\pi)^4}\,\frac{(1-\sigma)}{\sigma}\,
\left\{\,\delta(\tp-k_+)-\delta(\tp-(1-\sigma)k_+)\,\right\}\frac{1}{l^2(k-l)^2}\,,
\end{equation}
where the subscript $(c)$ denotes that this is the contribution
from the diagram of fig.~\ref{fig:da1loop}(c), and the superscript
$(1)$ that it is a one-loop contribution.
$\Phi^{b\bar{q}^\prime\,(1)}\otimes T^{(0)}\otimes
\Phi^{q\bar{q}^\prime\,(0)}$ is therefore
\begin{equation}
(3B)^\prime=2ig^2C_F\,\frac{1}{xk_+}\int
\frac{d^4l}{(2\pi)^4}\,\frac{1}{\sigma}\, \left\{\,(1-\sigma)
I_2(x,k_+)-I_2(x,(1-\sigma)k_+)\,\right\}\,\frac{1}{l^2(k-l)^2}\,.
\end{equation}
The results in section~\ref{subsubsec:o8collineark} are now
modified as follows:
\begin{eqnarray}
(23)&=&-2ig^2\left(C_F-\frac{C_A}{2}\right)\frac{1}{xk_+}\int\frac{d^4l}{(2\pi)^4}\,
\frac{I_2(x,(1-\sigma)k_+)}{\sigma}\,\frac{1}{l^2\,(k-l)^2}\,,\label{eq:collk23o1}\\
(34)&=&\ \ \,
2ig^2\left(C_F-\frac{C_A}{2}\right)\frac{1}{xk_+}\int\frac{d^4l}{(2\pi)^4}\,
\frac{1-\sigma}{\sigma}\,\frac{I_2(x,k_+)}{l^2\,(k-l)^2}\,,\label{eq:collk34o1}\\
(36)&=&-ig^2C_A\,\frac{1}{xk_+}\int\frac{d^4l}{(2\pi)^4}\,
\frac{I_2(x,k_+)}{l^2\,(k-l)^2}\,.
\end{eqnarray}
Finally we have to consider the diagrams (37) which no longer
vanish when we include the terms of $O(q_2\cdot k/m_b^2)$. Using
the Ward identity in eq.~(\ref{eq:wi0}) the contribution from
these diagrams is readily found to be:
\begin{equation}
(37)=ig^2C_A\,\frac{1}{xk_+}\int\frac{d^4l}{(2\pi)^4}\,
\frac{I_2(x,k_+)-I_2(x,(1-\sigma)k_+)}{\sigma}\,\frac{1}{l^2\,(k-l)^2}\,.
\end{equation}
Thus we have
\begin{equation}
(23)+(34)+(36)+(37)=(3B)^\prime\,,
\end{equation}
so that there are no mass singularities remaining in the
hard-scattering amplitude.

\section{Conclusions}\label{sec:concs}
In this paper, we have studied the radiative decays $B\to V\gamma$
(where $V=\rho, K^*$) in the framework of QCD factorisation. We
focused on spectator interactions, where the most significant
contributions come from the chromomagnetic operator ${\cal Q}_8$
and the four-quark operator ${\cal Q}_1$. These contributions had
previously been considered explicitly at leading
order~\cite{BB,BFS}, and they exhibit similar features to the
purely radiative decays $B\to\gamma\ell\nu$ ($\gamma\gamma$,
$\gamma\ell^+\ell^-$) that we have already
considered~\cite{dgs_glnu,dgs_universal}. However, the presence of
two light-cone distribution amplitudes (those of the $B$ and
vector mesons) leads to technical and conceptual modifications.

We have performed an explicit next-to-leading-order computation of
the contribution from ${\cal Q}_8$ in the heavy-quark limit
$m_b\to\infty$ (specifically we have calculated all the terms
containing mass singularities and large logarithms), and we showed
that the spectator interactions factorise, i.e. they can be
written as the convolution of a hard-scattering kernel, computable
in perturbation theory, and of two light-cone distribution
amplitudes (one for each meson) describing soft physics. The
explicit results are presented in appendix~\ref{app:explicit}. By
studying the soft and collinear regions of the loop momentum, we
can understand the factorisation of mass singularities using a
heuristic argument (presented in section~\ref{sec:fact1lpo8}). The
same argument can be applied to the 4-quark operator ${\cal Q}_1$,
so that its contribution to spectator interactions are also
expected to factorise at next-to-leading order. In
appendix~\ref{app:general} we use the collinear Ward identity to
demonstrate the factorisation of mass singularities at
next-to-leading order for all the weak operators contributing to
$\bvg$ decays.

This success of QCD factorisation for $B\to V\gamma$ spectator
interactions leaves several questions open. First, the presence of
an intermediate scale $\sqrt{\Lambda_{QCD} m_b}$, corresponding to
the virtuality of the exchanged gluon, yields large (Sudakov)
logarithms at all orders of perturbation theory. The latter could
be resummed using renormalization-group arguments within the
Soft-Collinear Effective Theory~\cite{SCET,SCET2,SCETvgam}; for
the resummation in purely radiative decays see
refs.~\cite{dgs_glnu,dgs_universal}. However, the mixing of the
operators and hence the subsequent resummation are much more
intricate than in the case of $b\to u$ transitions~\footnote{We
thank M.~Neubert for instructive discussions on this point.}.  We
have only considered twist-2 distribution amplitudes for both
mesons: higher-twist distribution amplitudes deserve a separate
study. One could also consider the interesting decay $B\to
V\ell^+\ell^-$, where the light vector meson is longitudinally
polarized and the lepton pair has a small invariant mass. In
ref.~\cite{BFS}, the spectator interactions were analysed at
leading order: they receive contributions from additional diagrams
to those considered here, involving (in particular) the second
leading-twist $B$-meson distribution amplitude $\Phi_-^B$. Of
course, one would also like to understand in detail hard spectator
interactions in two-body nonleptonic $B$-decays.

\section*{Acknowledgements}
We thank Gerhard Buchalla and Matthias Neubert for helpful
comments and discussion.

C.T.S. acknowledges partial support from PPARC through the grant
PPA/G/O/1998/00525. S.D.G. acknowledges partial support from
EU-RTN contract EURIDICE (HPRN-CT-2002-00311).

\appendix

\section{Contribution of ${\cal Q}_8$ at one loop} \label{app:explicit}

In this appendix we present the results from explicit evaluation
of the Feynman diagrams at one-loop order for the chromomagnetic
operator ${\cal Q}_8$. Specifically we exhibit the mass
singularities and all the large logarithms. We denote as $(ij)$
the contribution from the diagram where $i$ and $j$ are connected
with a gluon in Fig.~\ref{fig:1loop8}. Each diagram's contribution
is given in units of $\alpha_s/(4\pi){\cal A}_8^{(0)}$. We use
dimension regularisation to regulate both the ultraviolet and
infrared divergences, but introduce separate scales for the two
cases.

\subsection{Abelian Component}
We start by considering those diagrams which have an Abelian
component, i.e. which have a component proportional to $C_F$.

\subsubsection{Contributions to ${\cal A}_8^{(1)}$}

There are four diagrams with a gluon which links an initial-state
quark with a final-state one:
\begin{eqnarray}\label{eq:app12ab}
(12)&=& \left(C_F-\frac{C_A}{2}\right) \left[
    -\frac{1}{2}\log^2\frac{M_B^2}{\mir^2}
       -2\log\bar{x}\log\frac{M_B^2}{\mir^2}
 \right]\\
(14)&=& \left(C_F-\frac{C_A}{2}\right) \Bigg[
    \frac{1}{2}\log^2\frac{M_B^2}{\mir^2}
    +(2\log x-2)\log\frac{2(k\cdot q_2)}{\mir^2}
    -\log^2\frac{2(k\cdot q_2)}{M_B^2} \Bigg]\label{eq:app14ab}\\
(23)&=& \left(C_F-\frac{C_A}{2}\right) \left[
    \log^2\frac{2(k\cdot q_2)}{\mir^2}
    +2\log\bar{x}
       \log\frac{2(k\cdot q_2)}{\mir^2}
 \right]\label{eq:app23ab}\\
(34)&=& \left(C_F-\frac{C_A}{2}\right) \left[
    -\log^2\frac{2(k\cdot q_2)}{\mir^2}
    +(-2\log x+4)\log\frac{2(k\cdot q_2)}{\mir^2}
    -\log\frac{2(k\cdot q_2)}{\muv^2}
 \right]\!\!.\label{eq:app34ab}
\end{eqnarray}

One may also link the external lines of the same hadron, yielding
\begin{eqnarray*}
(13) &=& C_F
   \left[-\frac{1}{2}\log^2\left(-\left(\frac{2 k\cdot q_2}{M_B \mir}\right)^2
    \right)\right]\\
(24) &=& 0
\end{eqnarray*}
at leading twist.

When an external line is linked to the internal quark propagator
one gets
\begin{eqnarray*}
(15) &=& 0
\\
(25) &=& C_F
\left[2\log\frac{M_B^2}{\mir^2}-\log\frac{M_B^2}{\muv^2}
\right]\\
(35) &=& 0
\\
(45) &=& C_F \log\frac{2(k\cdot q_2)}{\mir^2}
   \left[2\frac{1}{x}\log\bar{x}+2\right]
\end{eqnarray*}

There are also contributions from the wave function
renormalization of the external lines
\begin{eqnarray*}
(11) &=& C_F \left[\frac{1}{2}\log\frac{M_B^2}{\muv^2}
   +\log\frac{M_B^2}{\mir^2}
\right]\\
(22) &=& 0 \quad\equiv\quad
   C_F \left[\frac{1}{2}\log\frac{\mir^2}{\muv^2}
\right]\\
(33) &=& 0 \quad\equiv\quad
   C_F \left[\frac{1}{2}\log\frac{\mir^2}{\muv^2}
\right]\\
(44) &=& 0 \quad\equiv\quad
   C_F \left[\frac{1}{2}\log\frac{\mir^2}{\muv^2}
\right]\\
(55) &=& C_F \left[\log\frac{M_B^2}{\muv^2} \right]
\end{eqnarray*}
where the vanishing values correspond to massless tadpole diagrams
in dimensional regularisation.

\begin{figure}[t]
\begin{center}
\begin{picture}(170,70)(-15,-5)
\SetWidth{1.5}\Line(0,60)(40,60)\SetWidth{0.8}
\ArrowLine(9,60)(10,60)\SetWidth{0.5}
\PhotonArc(70,60)(30,0,180){2}{16}
\Line(40,60)(140,60)\ArrowLine(130,60)(131,60)
\GBox(35,55)(45,65){1}\Photon(90,60)(90,0){2}{12}
\Gluon(50,60)(70,40){2}{3} \Line(0,0)(140,0)\ArrowLine(9,0)(10,0)
\ArrowLine(130,0)(131,0)\ArrowLine(70,40)(71,39)
\Text(-5,60)[r]{\small{$p$\,--\,$k$}} \Text(-5,0)[r]{\small{$k$}}
\Text(145,60)[l]{\small{$\bar{x}q_2$}}
\Text(145,0)[l]{\small{$xq_2$}} \Text(70,35)[t]{\small{$q_1$}}
\end{picture}
\caption{An additional diagram, $D$, which contains a large
logarithm and which is not included in the set of diagrams
\{\,$(ij)\,$\} defined in section~\ref{sec:fact1lpo8}.
\label{fig:diagramD}}
\end{center}
\end{figure}
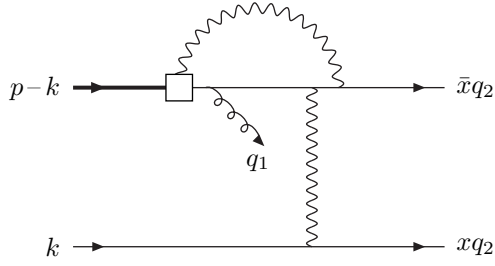

Finally, there is a contribution containing large logarithms that
cannot be obtained by adding one gluon to the leading-twist
tree-level diagram and is therefore not included in the set
\{\,$(ij)$\,\} defined in section~\ref{sec:fact1lpo8}. We denote
it by $D$ and exhibit it in fig.~\ref{fig:diagramD}; its
contribution to the amplitude is:
\begin{equation}
D=(C_F-C_A/2)\frac{\bar{x}}{x}
   \left[-2\log\left(-\frac{2k\cdot q_2}{M_B^2}\right)
         -\log\left(-\frac{M_B^2}{\muv^2}\right)\right]\,.
\end{equation}
The other diagrams not included in the set \{\,$(ij)$\,\} do not
give large logarithms when $\muv$ is set equal to $M_B$.

\subsubsection{Distribution amplitudes}

We consider now the contribution to the distribution amplitudes
convoluted with the lowest-order kernel, $\Phi^{b\bar{q}'\,(1)}
\otimes T_8^{(0)} \otimes \Phi^{q\bar{q}'\,(0)}+
\Phi^{b\bar{q}'\,(0)} \otimes T_8^{(0)} \otimes
\Phi^{q\bar{q}'\,(1)}$\,. As explained in
section~\ref{sec:fact1lpo8}, we denote the diagrams in a similar
way to those for the amplitude. Once again we use dimensional
regularisation to regulate both infared and ultraviolet
divergences. The infrared scale, $\mir$, is the same for the
matrix elements and the distribution amplitudes. The ultraviolet
scale is different in the two cases however: for the matrix
elements $\muv$ corresponds to the renormalization scale $\mu_R$,
whereas for the distribution amplitudes $\nuv$ corresponds to the
factorisation scale $\mu_F$ (see ref.~\cite{dgs_glnu} for further
discussion). In agreement with our definition of $\Phi^H$, we use
the HQET Lagrangian when a gluon is connected to the $b$-quark.

We can link an external line to the path-ordered exponential of
one distribution amplitude
\begin{eqnarray*}
(1B)' &=& C_F \left[-\frac{1}{2}\log^2\frac{2k_+^2}{\nuv^2}
\right]\\
(2V)' &=& 0 \equiv C_F \left[2\log\frac{\nuv^2}{\mir^2}
\right]\\
(3B)' &=& 0 \equiv C_F \left[2\log\frac{\nuv^2}{\mir^2}
\right]\\
(4V)' &=&
 C_F \log\frac{\nuv^2}{\mir^2}
\left[2\frac{1}{x}\log\bar{x}+2\right]\,.
\end{eqnarray*}

We can also link two external lines
\begin{eqnarray*}
(13)' &=& C_F
   \left[-\frac{1}{2}\log^2\left(-\left(\frac{2 k\cdot q_2}{M_B \mir}\right)^2
    \right)\right]
\\
(24)' &=& 0
\end{eqnarray*}

There are also contributions due to the wave function
renormalization for external lines
\begin{eqnarray*}
(11)' &=& C_F \left[\log\frac{\nuv^2}{\mir^2}
\right]\\
(22)' &=& 0 \quad\equiv\quad
   C_F \left[\frac{1}{2}\log\frac{\mir^2}{\nuv^2}
\right]\\
(33)' &=& 0 \quad\equiv\quad
   C_F \left[\frac{1}{2}\log\frac{\mir^2}{\nuv^2}
\right]\\
(44)' &=& 0 \quad\equiv\quad
   C_F \left[\frac{1}{2}\log\frac{\mir^2}{\nuv^2}
\right]
\end{eqnarray*}
where the vanishing values correspond to massless tadpole diagrams
in dimensional regularisation.

We could also try to link two points of the same path-ordered
exponential. These diagrams involve the propagator associated with
$A^+(\alpha z) A^+(\beta z)$ or $A^-(\alpha z) A^-(\beta z)$,
which vanishes in the Feynman gauge.

\subsection{Remaining Nonabelian Components}

We now evaluate the one-loop contributions to ${\cal A}_8^{(1)}$
whose colour factor is proportional to $C_A$.

\subsubsection{Diagrams with one Three-Gluon Vertex}

The contributions from diagrams in which a gluon links an external
quark line to the gluon one through a three-gluon vertex are as
follows:
\begin{eqnarray*}
(16) &=& C_A \left[-\frac{1}{2}\log^2 \frac{2(k\cdot q_2)}{M_B^2}
  +\left(-\frac{5}{4}+\log x\right)\log \frac{2(k\cdot q_2)}{M_B^2}
  -\left(\frac{1}{2x}+2\right)\log \frac{M_B^2}{\muv^2}
\right]\\
(26) &=& C_A \times \frac{\bar{x}}{x}
  \left[\frac{1}{2}\left(1-\frac{\log\bar{x}}{x}\right)
   \log \frac{2(k\cdot q_2)}{M_B^2}
  -\log \frac{2(k\cdot q_2)}{\mir^2}
  +\frac{1}{8}\log \frac{M_B^2}{\muv^2}
\right]\,,\\
(36) &=& C_A \left[-\frac{3}{2}\log\frac{2(k\cdot q_2)}{\muv^2}
                   +2\log\frac{2(k\cdot q_2)}{\mir^2}
\right]\,,
\end{eqnarray*}
keeping in mind that $(46)$ is the same diagram as $(36)$.

In the case of the internal quark propagator, we get
$$
(56)= C_A \left[
\left(-\frac{1}{2x}-\frac{3}{4}-\frac{1+x}{2x^2}\log\bar{x}\right)
   \log\frac{2(k\cdot q_2)}{M_B^2}
 +\left(-\frac{5}{8x}-\frac{11}{8}\right)\log\frac{M_B^2}{\muv^2}
\right]\,.
$$

\subsubsection{Diagrams for the Gluonic Vacuum Polarization}

The gluonic wave function renormalization yields
$$
(66)= \left[\frac{2}{3}N_f -\frac{5}{3}C_A \right]
 \log\frac{2(k\cdot q_2)}{\muv^2}\,.
$$

\subsubsection{Diagrams with two Gluons attached to ${\cal Q}_8$}

We can also attach two gluons to the vertex with the operator
${\cal Q}_8$. One of them interacts with the spectator quark,
whereas the second is attached elsewhere.

When one of the gluons is attached to an external line, we obtain
\begin{eqnarray*}
(17) &=& C_A \times \frac{5}{4x}\log\frac{M_B^2}{\muv^2}
\\
(27) &=& C_A \times
\frac{\bar{x}}{x}\left[-\frac{1}{2}\log\frac{M_B^2}{\muv^2}
  +\log\frac{M_B^2}{\mir^2}
\right]\\
(37) &=& C_A \times (-1)\log\frac{2(k\cdot q_2)}{\mir^2}
\end{eqnarray*}
where $(37)$ and $(47)$ denote the same diagram.

The result for diagram (57) obtained by attaching one of the
gluons to the internal quark propagator is
$$
(57)= C_A \times \frac{\bar{x}}{4x}\log\frac{M_B^2}{\muv^2}
$$
and for diagram (67) in which the gluon is attached the internal
gluon propagator is
$$
(67)= C_A \times \frac32 \log\frac{2(k\cdot q_2)}{\muv^2}\,.
$$

\subsection{One-loop results}

We now combine the results presented above to obtain the large
logarithms at NLO in the amplitude, in $\Phi^B\otimes T_8\otimes
\Phi_\perp$ and hence finally in the hard-scattering kernel. These
next-to-leading-order results contain logarithms of ratios of the
scales $\mu_R$, $\mu_F$, $M_B$ and $q_2\cdot k$. Following
ref.~\cite{dgs_glnu,dgs_universal}, the natural choices for the
renormalisation and factorisation scales are
\begin{equation}
\mu_R= O(M_B) \qquad \mu_F= \mu_i = O(q_2\cdot k)
\end{equation}
But even with this choice, large logarithms will remain due to the
presence of three distinct scales. Therefore, in presenting our
results we focus on the mass singularities and large logarithms.
The amplitude up to one-loop order is:
\begin{eqnarray}
{\mathcal{A}}_8^{(0+1)}(\mu_R=M) &=&
  {\mathcal{A}}_8^{(0)}(\mu_R=\mu_i)\nonumber\\
&&\hspace{-1.2in}\times \Bigg[1+\frac{\as}{4\pi}
  \Bigg\{
    C_F\Bigg[
          \frac{7}{2}\log\frac{\mu_i^2}{\mir^2}
          +\left(\frac{2}{x}\log\bar{x}+2\right)\log\frac{\mu_i^2}{\mir^2}
          -\frac{1}{2}\log\left(-\left(\frac{\mu_i^2}{M_B \mir}\right)^2\right)
          -\log^2\frac{\mu_i^2}{M_B^2}
\nonumber\\
&&\hspace{-1.3in}
          +\left[2\log\bar{x}-2\frac{\bar{x}}{x}-\frac{5}{2}\right]
              \log\frac{\mu_i^2}{M_B^2}
       \Bigg]+C_A\left[-\log\frac{\bar{x}}{x}-\frac{1}{x^2}\log\bar{x}\right]
        \log\frac{\mu_i^2}{M_B^2}
  +\ldots   \Bigg\}
\Bigg] + O(\as^2)\label{eq:a81oneloop}
\end{eqnarray}
where ${\mathcal{A}}_8^{(0)}(\mu_R)$
denotes the tree-level matrix element with $\mu_R$ as the scale of the strong
coupling constant, and the ellipses denote terms without large logarithms.

The contribution from the distribution amplitudes is
\begin{eqnarray} \label{eq:phi81oneloop}
[\Phi^H \otimes T_8 \otimes \Phi^V]^{(0+1)}(\mu_F=\mu_i)&=&
  {\mathcal{A}}_8^{(0)}\times\Bigg[1+\frac{\as}{4\pi}C_F\times\\
&&\hspace{-2.2in}
    \Bigg[
          \frac{7}{2}\log\frac{\mu_i^2}{\mir^2}
          +\left(\frac{2}{x}\log\bar{x}+2\right)\log\frac{\mu_i^2}{\mir^2}
          -\frac{1}{2}\log\left(-\left(\frac{\mu_i^2}{M_B \mir}\right)^2\right)
 -\frac{1}{2}\log^2\frac{\mu_i^2}{M_B^2}
       \Bigg]\Bigg]  + O(\as^2)\,. \nonumber
\end{eqnarray}

The one-loop result for the hard-scattering kernel at
next-to-leading order is therefore:
\begin{eqnarray} \label{eq:t81oneloop}
T_8^{(0+1)}(\tilde{k}_+,u;\mu_R=M,\mu_F=\mu_i) &=&
  T_8^{(0)}(\tilde{k}_+,u;\mu_R=\mu_i)\times\Bigg[1+\frac{\as}{4\pi}\times
\\
&&\hspace{-2.5in}
  \Bigg\{-\frac{C_F}{2}\log^2\frac{\mu_i^2}{M_B^2}
+C_F\left[2\log\bar{x}-2\frac{\bar{x}}{x}-\frac{5}{2}\right]
              \log\frac{\mu_i^2}{M_B^2}+C_A\left[-\log\frac{\bar{x}}{x}
              -\frac{1}{x^2}\log\bar{x}\right]
        \log\frac{\mu_i^2}{M_B^2}
  +\ldots   \Bigg\}
\Bigg]\nonumber\\&& + O(\as^2) \nonumber
\end{eqnarray}
where $T_8^{(0)}(\tilde{k}_+,u;\mu_R)$ denotes the tree-level
hard-scattering kernel with $\mu_R$ as the scale of the strong
coupling constant, and the ellipses denote terms without large
logarithms. The key point is, of course, that there are no mass
singularities in the expression for $T_8$.

\section{Cancellation of Collinear Divergences and the Collinear
Ward Identity}\label{app:general}

In this appendix we show that the cancellations of collinear
divergences demonstrated in sections~\ref{sec:fact1lpo8} and
\ref{sec:fact1lpo1} are a general consequence of the collinear
Ward identity and hold for all the weak operators contributing to
$\bvg$ decays~\footnote{Similar arguments were presented in a
different context in Ref.~\cite{Li}.}. It will be useful to
consider (in perturbation theory) the on-shell amplitude
$b\bar{q}^\prime\to q\bar{q}^\prime\gamma\,+$\,gluon, where the
momentum of the gluon is $l$. We represent this amplitude by the
graphs in Fig.~\ref{fig:wi1}, exhibiting explicitly the diagrams
in which the gluon is attached to an external quark line. We write
this amplitude as $\varepsilon^\mu_g(l) {\cal M}_\mu$, where
$\varepsilon_g$ is the polarization vector of the gluon. The Ward
Identity we wish to exploit is
\begin{equation}\label{eq:wi}
l^\mu\,{\cal M}_\mu=0\,.
\end{equation}

\begin{figure}
\begin{center}
\begin{picture}(430,210)(0,-150)
\SetWidth{1.5}\ArrowLine(0,25)(40,15)\SetWidth{0.5}
\ArrowLine(0,-25)(40,-15)\ArrowLine(60,15)(100,25)
\ArrowLine(60,-15)(100,-25)
\Gluon(70,0)(100,0){2}{4}\ArrowLine(99,0)(100,0)
\GCirc(50,0){20}{0.9}\Photon(10,22.5)(40,50){2}{5}
\ArrowLine(40,50)(41,51)\Text(46,50)[l]{\boldmath{$l$}}
\Text(95,-5)[tl]{\boldmath{$q_1$}}
\Text(130,0)[]{+}
\SetWidth{1.5}\ArrowLine(160,25)(200,15)\SetWidth{0.5}
\ArrowLine(160,-25)(200,-15)\ArrowLine(220,15)(260,25)
\ArrowLine(220,-15)(260,-25)\Gluon(230,0)(260,0){2}{4}
\GCirc(210,0){20}{0.9}\Photon(170,-22.5)(200,-40){2}{5}
\Text(290,0)[]{+}
\SetWidth{1.5}\ArrowLine(320,25)(360,15)\SetWidth{0.5}
\ArrowLine(320,-25)(360,-15)\ArrowLine(380,15)(420,25)
\ArrowLine(380,-15)(420,-25)\Gluon(390,0)(420,0){2}{4}
\GCirc(370,0){20}{0.9}\Photon(405,21.25)(430,45){2}{5}
\Text(60,-100)[]{+}
\SetWidth{1.5}\ArrowLine(80,-75)(120,-85)\SetWidth{0.5}
\ArrowLine(80,-125)(120,-115)\ArrowLine(140,-85)(180,-75)
\ArrowLine(140,-115)(180,-125)\Gluon(150,-100)(180,-100){2}{4}
\GCirc(130,-100){20}{0.9}\Photon(165,-121.25)(185,-145){2}{5}
\Text(210,-100)[]{+}
\SetWidth{1.5}\ArrowLine(240,-75)(280,-85)\SetWidth{0.5}
\ArrowLine(240,-125)(280,-115)\ArrowLine(300,-85)(340,-75)
\ArrowLine(300,-115)(340,-125)\Gluon(310,-95)(340,-95){2}{4}
\Photon(310,-105)(335,-110){2}{4} \GCirc(290,-100){20}{0.9}
\end{picture}
\end{center}
\caption{Amplitude for the auxiliary process $b\bar{q}\to
q\bar{q}\gamma$ + gluon.\label{fig:wi1}. The spring represents the
photon (with momentum $q_1$) and the curly line the gluon (with
outgoing momentum $l$). The first four diagrams are those in which
the final-state gluon is attached to one of the four external
quark lines. The fifth diagram represents all the remaining ways
in which the final-state gluon can be attached.}
\end{figure}
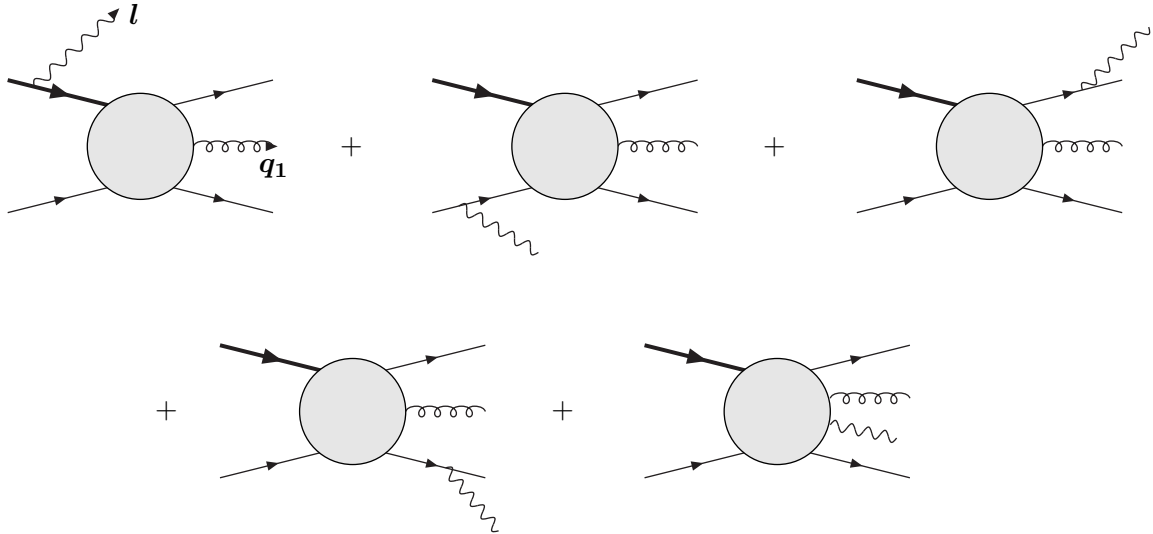

\begin{figure}
\begin{center}
\begin{picture}(430,210)(0,-150)
\Text(0,0)[]{$l^\mu\ \times$}
\SetWidth{1.5}\ArrowLine(0,25)(40,15)\SetWidth{0.5}
\ArrowLine(0,-25)(40,-15)\ArrowLine(60,15)(100,25)
\ArrowLine(60,-15)(100,-25)\Gluon(70,0)(100,0){2}{4}
\GCirc(50,0){20}{0.9}\Photon(10,22.5)(40,50){2}{5}
\ArrowLine(40,50)(41,51)\Text(46,50)[l]{\boldmath{$l$}}
\Text(95,-5)[tl]{\boldmath{$q_1$}}
\Text(130,0)[]{+}\Text(160,0)[]{$l^\mu\ \times$}
\SetWidth{1.5}\ArrowLine(160,25)(200,15)\SetWidth{0.5}
\ArrowLine(160,-25)(200,-15)\ArrowLine(220,15)(260,25)
\ArrowLine(220,-15)(260,-25)\Gluon(230,0)(260,0){2}{4}
\GCirc(210,0){20}{0.9}\Photon(170,-22.5)(200,-40){2}{5}
\Text(290,0)[]{+}\Text(320,0)[]{$l^\mu\ \times$}
\SetWidth{1.5}\ArrowLine(320,25)(360,15)\SetWidth{0.5}
\ArrowLine(320,-25)(360,-15)\ArrowLine(380,15)(420,25)
\ArrowLine(380,-15)(420,-25)\Gluon(390,5)(420,5){2}{4}
\Photon(390,-5)(415,-10){2}{4}\GCirc(370,0){20}{0.9}
\Text(40,-100)[]{=}\Text(70,-100)[]{$-\,i\ \times$}
\SetWidth{1.5}\ArrowLine(80,-75)(120,-85)\SetWidth{0.5}
\ArrowLine(80,-125)(120,-115)\ArrowLine(140,-85)(180,-75)
\ArrowLine(140,-115)(180,-125)\Gluon(150,-100)(180,-100){2}{4}
\GCirc(130,-100){20}{0.9}\Photon(146,-83.5)(175,-55){2}{5}
\GCirc(146,-83.5){2}{0}\Text(210,-100)[]{+}\Text(230,-100)[l]{$i\
\times$} \SetWidth{1.5}\ArrowLine(240,-75)(280,-85)\SetWidth{0.5}
\ArrowLine(240,-125)(280,-115)\ArrowLine(300,-85)(340,-75)
\ArrowLine(300,-115)(340,-125)\Gluon(310,-95)(340,-95){2}{4}
\GCirc(290,-100){20}{0.9}\Photon(306,-116.5)(335,-145){2}{6}
\GCirc(306,-116.5){2}{0}
\end{picture}
\end{center}
\caption{\label{fig:wi2} Representation of the Ward identity in
eq.~(\ref{eq:wi}). The solid dot in the two diagrams on the
right-hand side indicates that the gluon is attached with factor
$igT^a$ (but no $\gamma$-matrix) and the propagator to the left of
the dot has been eliminated using eq.~(\ref{eq:wi0}).}
\end{figure}
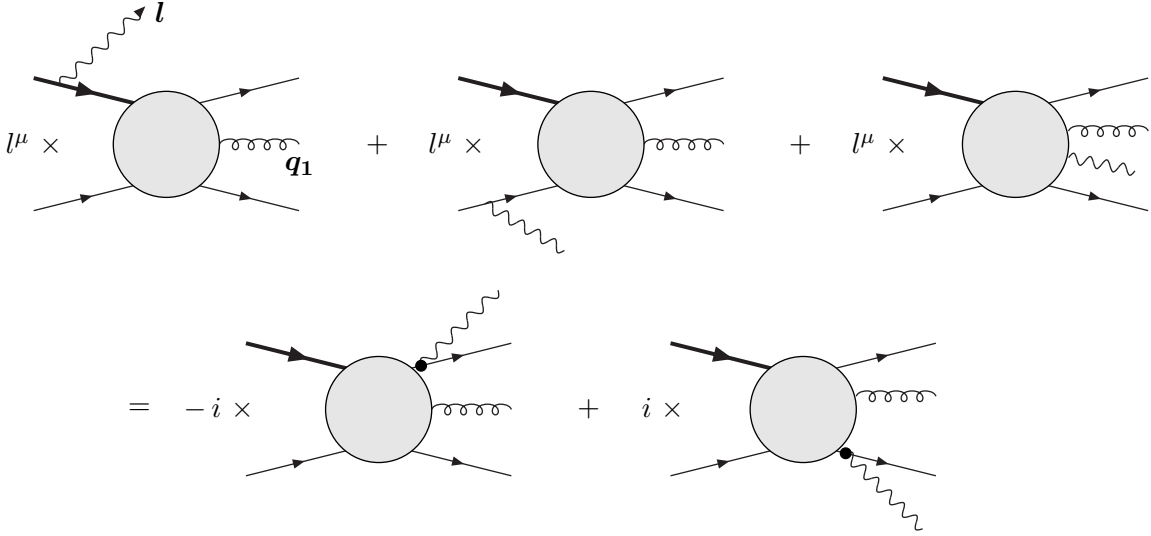

The process we are actually studying in the evaluation of the
one-loop contributions to the hard-scattering kernels is
$b\bar{q}^\prime\to q\bar{q}^\prime\gamma$, where the momenta of
the incoming $b$-quark and light antiquark are $q_1+q_2-k$ and $k$
respectively and those of the outgoing quark and antiquark are
$\bar{x}q_2$ and $xq_2$. We denote this amplitude at lowest order
in perturbation theory by ${\cal A}^{(0)}(x,k)$, without
specifying which operator mediates the weak transition. The
corresponding hard-scattering kernel is written as $T^{(0)}$,
where we suppress the spinor indices.

We start by the demonstration of the cancellation of collinear
divergences from the region in which $l$ is collinear to $q_2$,
$l\simeq\xi q_2$. It is convenient to use the identity in
eq.~(\ref{eq:wi0}) to represent the Ward identity (\ref{eq:wi}) by
Fig.~\ref{fig:wi2}. On the right-hand side we have eliminated the
fermion propagator immediately to the left of the solid dot. Since
the external particles are on-shell we can use the identity in
Fig.~\ref{fig:wi2} with the external wave functions amputated.

We now consider the identity in Fig.~\ref{fig:wi2} with the
external wave functions amputated, and with the momentum of the
final state quark equal to $\bar{x}q_2-l$. We multiply both sides
of Fig.~\ref{fig:wi2} by the factor
\begin{equation}\label{eq:factor}
-2ig\,\frac{\bar{x}-\xi}{\xi}\,\frac{1}{l^2(\bar{x}q_2-l)^2}\,,
\end{equation}
and the outgoing quark line by the colour matrix $T^a$ ($a$ is the
colour label of the final-state gluon). Integrating over $l$ (in
the collinear region), on the left hand side we generate all the
diagrams with a gluon attached to final-state quark line (i.e. the
line denoted by 2), except for (22) and (24)~\footnote{Apart from
the factor in eq.~(\ref{eq:factor}), we have set
$(\bar{x}q_2-l)^2=0$ everywhere, which is correct at leading
twist.}. From the identity in Fig.~\ref{fig:wi2}, we then
immediately understand that for any operator the sum of these
diagrams gives a contribution which can be written in terms of a
convolution of the properties of the final state with the
lowest-order amplitude. Specifically, Fig~\ref{fig:wi2} gives
\begin{eqnarray}
\sum_{n\neq
2,4}(2n)&=&2ig^2C_F\int\frac{d^4l}{(2\pi)^4}\,\frac{\bar{x}-\xi}{\xi}\,
\frac{{\cal A}^{(0)}(x,k)-{\cal A}^{(0)}(x+\xi,k)}{l^2(\bar{x}q_2-l)^2}\,\\
&=&\Phi^{b\bar{q}^\prime\,(0)}\otimes T^{(0)}\otimes
\Phi_{(a)}^{q\bar{q}^\prime\,(1)}\\
&=&(2V)^\prime\,,
\end{eqnarray}
where the second line follows from eq.~(\ref{eq:phi1a}). Here we
have restored the external quark wave functions.

Following similar arguments and eq.~(\ref{eq:phi1b}) one also
obtains
\begin{eqnarray}
\sum_{n\neq
2,4}(4n)&=&2ig^2C_F\int\frac{d^4l}{(2\pi)^4}\,\frac{x-\xi}{\xi}\,
\frac{{\cal A}^{(0)}(x,k)-{\cal A}^{(0)}(x-\xi,k)}{l^2(xq_2-l)^2}\,\\
&=&\Phi^{b\bar{q}^\prime\,(0)}\otimes T^{(0)}\otimes
\Phi_{(b)}^{q\bar{q}^\prime\,(1)}\\
&=&(4V)^\prime\,.
\end{eqnarray}

Finally one can use the same procedure for the collinear
divergences from the region in which $l$ is parallel to $k$
($l\simeq\sigma k$). In this case we use the identity in
eq.~(\ref{eq:wi0}) to eliminate the propagators to the right of
the gluon vertex in the first two diagrams of Fig.~\ref{fig:wi1},
and obtain
\begin{eqnarray}
\sum_{n\neq
1,3}(3n)&=&2ig^2C_F\int\frac{d^4l}{(2\pi)^4}\,\frac{1-\sigma}{\sigma}\,
\frac{{\cal A}^{(0)}(x,k)-{\cal A}^{(0)}(x,(1-\sigma)k)}{l^2(k-l)^2}\,\\
&=&\Phi_{(c)}^{b\bar{q}^\prime\,(1)}\otimes T^{(0)}\otimes \Phi^{q\bar{q}^\prime\,(0)}\\
&=&(3B)^\prime\,.
\end{eqnarray}

Thus all the collinear divergences are absorbed into the
light-cone distribution amplitudes for any of the weak operators
${\cal Q}_i$ (the argument presented here considers all
singularities from the collinear region, and includes in
particular the soft-collinear modes discussed recently in
refs.~\cite{SCmodes}). It is straightforward to verify that for
${\cal Q}_8$, for which
\[A^{(0)}(x,k)\propto\frac{1}{\bar{x}k_+}\,,\] one recovers the
results in section~\ref{sec:fact1lpo8} and similarly for ${\cal
Q}_1$ one recovers the results in \ref{sec:fact1lpo1}.

\end{document}